\documentclass[prl,aps,twocolumn,floats,floatfix,superscriptaddress,reprint]{revtex4-1}
\usepackage{graphicx}
\usepackage{amsmath}
\usepackage{amssymb}
\usepackage{amsfonts}
\usepackage{times}
\def\nin{\noindent}
\def\non{\nonumber}
\def\be{\begin{equation}}
\def\ee{\end{equation}}
\def\bea{\begin{eqnarray}}
\def\eea{\end{eqnarray}}
\def\r{{\bf r}}
\def\k{{\bf k}}

\def\D{{\cal D}}

\def\O{{\cal O}}

\def\lp{\left(}
\def\rp{\right)}
\def\lb{\left[}
\def\rb{\right]}
\def\la{\left<}
\def\ra{\right>}

\def\Dinf{D_{\infty}}

\def\w{\omega}
\def\d{\mbox{d}}

\def\Re{\mbox{\,Re\,}}

\def\tp{{\tilde p}}

\def\alpha{\vartheta}

\begin{document}

%\bibliographystyle{plainnat}
%\bibpunct{(}{)}{,}{n}{}{,} % Science
%\renewcommand{\bibnumfmt}[1]{{#1}.} % Science
%\renewcommand{\bibfont}{\small}
%\renewcommand{\bibnumfmt}[1]{\textbf{#1}:}

\title{Characterizing microstructure of living tissues with time-dependent diffusion}
%\title{\Large Quantifying microstructure with time-dependent diffusion, from muscles to brain}
%\title{\Large Classifying and characterizing microstructure with time-dependent diffusion}

\author{Dmitry S. Novikov}
\email{dima@alum.mit.edu}
\affiliation{Bernard and Irene Schwartz Center for Biomedical Imaging, Department of Radiology,
New York University School of Medicine, New York, NY 10016, USA}
\author{Els Fieremans}
\affiliation{Bernard and Irene Schwartz Center for Biomedical Imaging, Department of Radiology,
New York University School of Medicine, New York, NY 10016, USA}
\author{Jens H. Jensen}
\affiliation{Department of Radiology and Radiological Science, Medical University of South Carolina, Charleston, SC 29425, USA}
%\affiliation{Department of Physiology and Neuroscience,
%New York University School of Medicine, New York, New York, USA}
\author{Joseph A. Helpern}
\affiliation{Department of Radiology and Radiological Science, Medical University of South Carolina, Charleston, SC 29425, USA}

\date{\today}
%\maketitle

\begin{abstract}
\nin
{
Molecular diffusion measurements are widely used to probe microstructure in materials and living organisms noninvasively. The precise relation of diffusion metrics to microstructure remains a major challenge: In complex samples, it is often unclear which structural features are most relevant and can be quantified. Here we classify the structural complexity in terms of the long time tail exponent in the molecular velocity autocorrelation function. The specific values of the dynamical exponent let us identify the relevant tissue microanatomy affecting water diffusion measured with MRI in muscles and in brain, and the microstructural changes in ischemic stroke. Our framework presents a systematic way to identify the most relevant part of structural complexity using transport measured with a variety of techniques.
}
\end{abstract}

\maketitle

When modeling or interpreting bulk transport in realistic disordered samples, e.g. composites, porous rocks, or living tissues, the challenge is to identify what part of microstructure to focus on
\cite{HausKehr,Bouchaud,callaghan,kusumi2005,Mitra92,latour-pnas,LeBihan,mair-prl1999}. 
Macroscopic transport is affected by multiple aspects of the immense microscopic complexity,
yet their relative importance is hard to estimate and compare.
%Every aspect of the immense microscopic complexity affects macroscopic transport properties, yet its relative importance is hard to estimate reliably. 
Here, we describe how a measurement itself may give us a hint regarding which parts of the microstructure are most relevant and, thereby, can be quantified. To that end, we suggest to employ
%In this work we focus
the long temporal correlations in molecular diffusion which preserve the footprint of the underlying structural complexity.
%and apply our framework to identify the relevant anatomical features in time-dependent diffusion-weighted magnetic resonance imaging (dMRI) measurements of water diffusion in muscles  \cite{kim-mrm2005} and in cerebral gray matter  \cite{gore2003}.  
These correlations manifest themselves in the power law tail of the molecular velocity autocorrelation function
\be \label{vaf}
\D(t) \equiv  \langle v(t) v(0) \rangle \sim t^{-(1+\alpha )} \,, \quad \alpha>0 \,. 
\ee
Practically, the tail \eqref{vaf} can be identified in the way the time-dependent instantaneous diffusion coefficient 
\be \label{Dinst-main}
D_{\rm inst}(t)\equiv {\partial \over \partial t} {\langle \delta x^{2}\rangle \over 2}
 = \int_{0}^{t}\! \d t' \, \D(t') \simeq \Dinf + \mbox{const}\cdot t^{-\alpha}
\ee
approaches the finite bulk diffusion constant $\Dinf$. The quantity $D_{\rm inst}(t)$ is   
accessible with any technique   \cite{callaghan,kusumi2005} measuring the mean square molecular displacement $\langle \delta x^{2}(t)\rangle$ in a particular direction, 
 see Eqs.~\eqref{Dcum}--\eqref{Dw} in the Supplemental Material \cite{supp}.

\begin{figure}[t]
\flushleft{\bf A}\\
\includegraphics[width=3.1in]{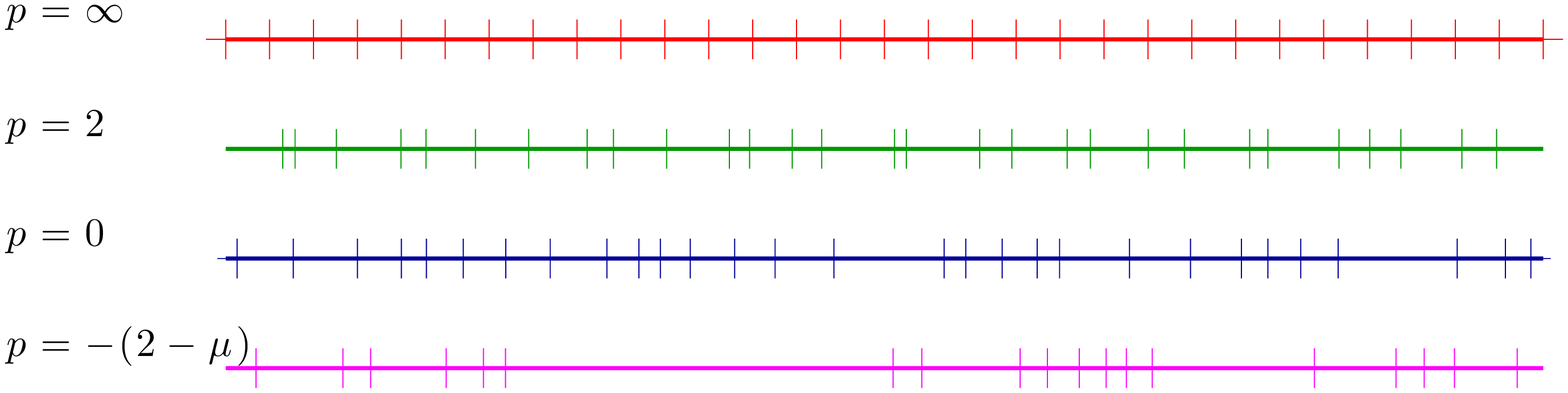}%
\\ %\vspace{0.1in}
{\bf B}\\
\includegraphics[width=3.1in]{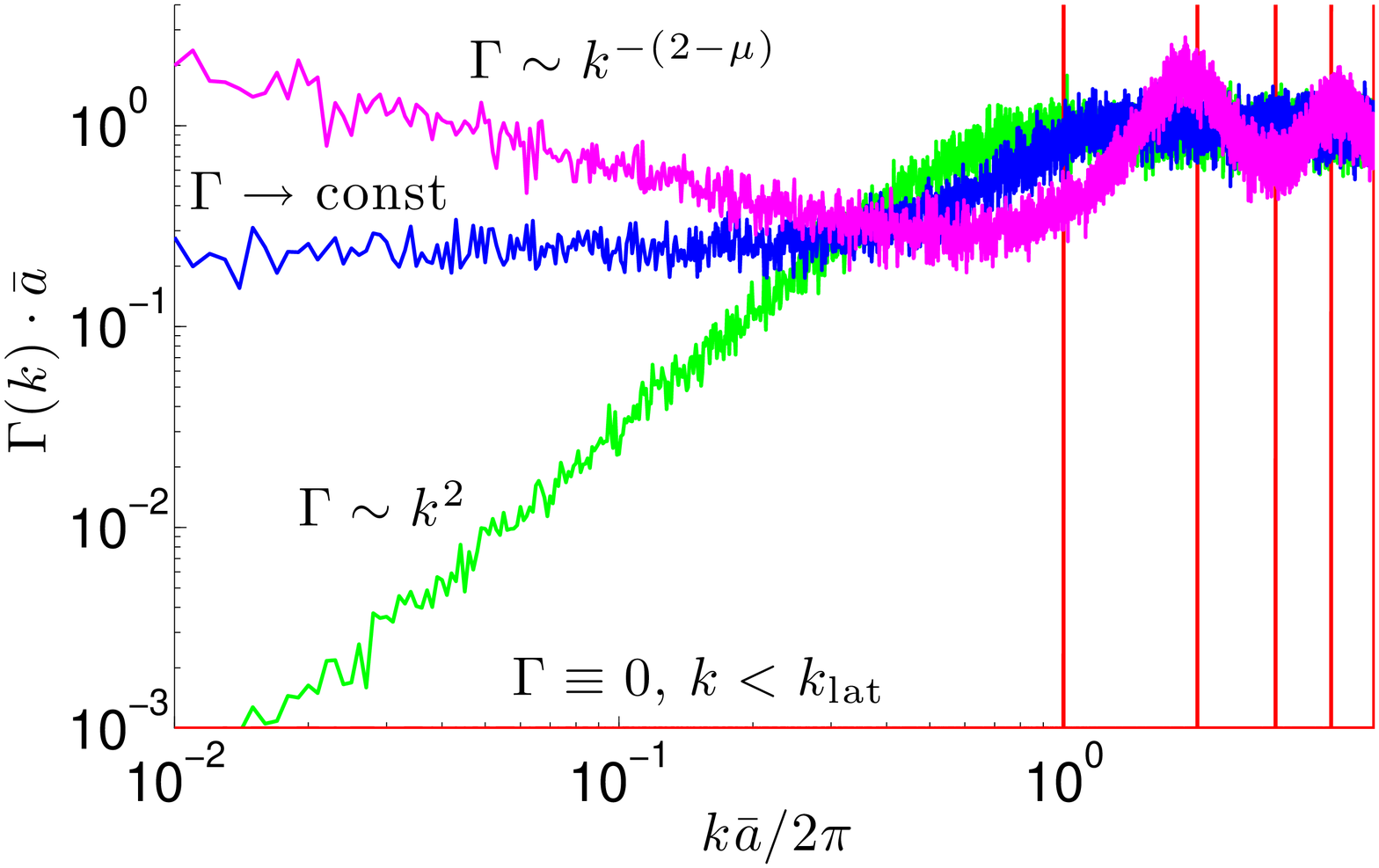}%
\\ \vspace{-0.1in}
{\bf C}\\
\includegraphics[width=3.1in]{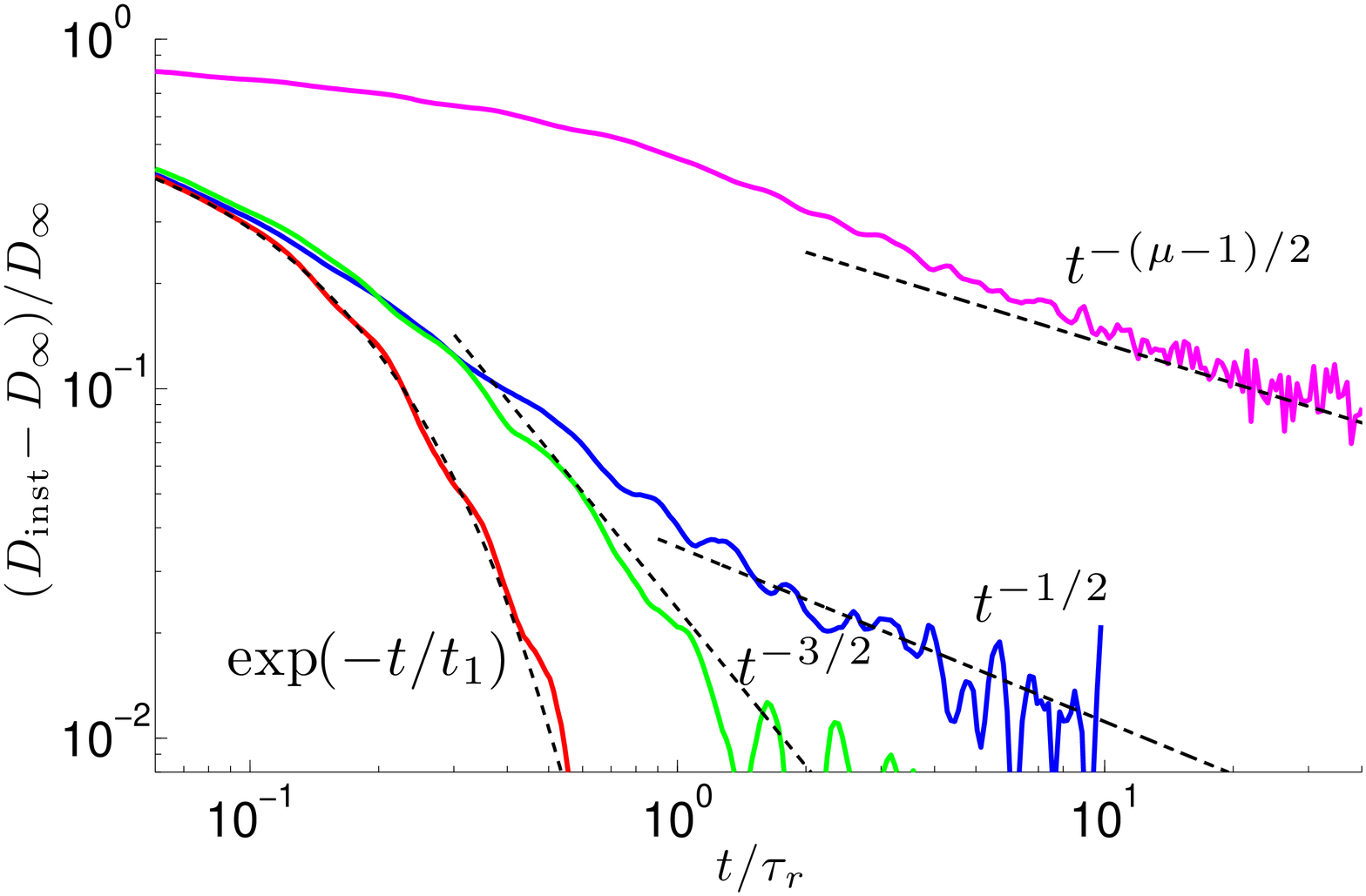}%
\caption{Time-dependent diffusion distinguishes between structural complexity classes in one dimension, represented by the placement of identical permeable barriers with the same mean density.  
{\bf (A)}  {\it Order} (red),
{\it hyperuniform disorder} (green), 
{\it short-range disorder} (blue), 
and {\it strong disorder} (magenta).
% with the corresponding exponents $p$ in the barrier density correlator $\Gamma(k)\sim k^{p}$. 
{\bf (B)} The barrier densities have qualitatively different large-scale fluctuations, reflected in the small-$k$ behavior of  their density correlator  $\Gamma(k)\sim k^{p}$ (see text). 
%for the short-range and $p=2$ for the hyperuniform disorder. For the ordered barriers, $\Gamma\equiv 0$ for small $k$, which can be formally represented by $p=\infty$.  
{\bf (C)} Numerical results confirming the relation \eqref{alpha=p+d}. 
%For the ordered case, $D_{\rm inst}(t)-\Dinf$ decreases exponentially ($\alpha=\infty$).
The time-dependence \eqref{Dinst-main} clearly distinguishes between the four arrangements, while the value $\Dinf$ is the same for all of them. 
The dashed lines are the exact power laws from equation \eqref{dD-corr}, and the exponential decrease is from the exact solution (see text and \cite{supp}); $\tau_{r}=\bar a/2\kappa$.
}
\label{fig:1d}
\end{figure}

\begin{figure}[t]
\flushleft
{\bf A}\includegraphics[width=1.5in]{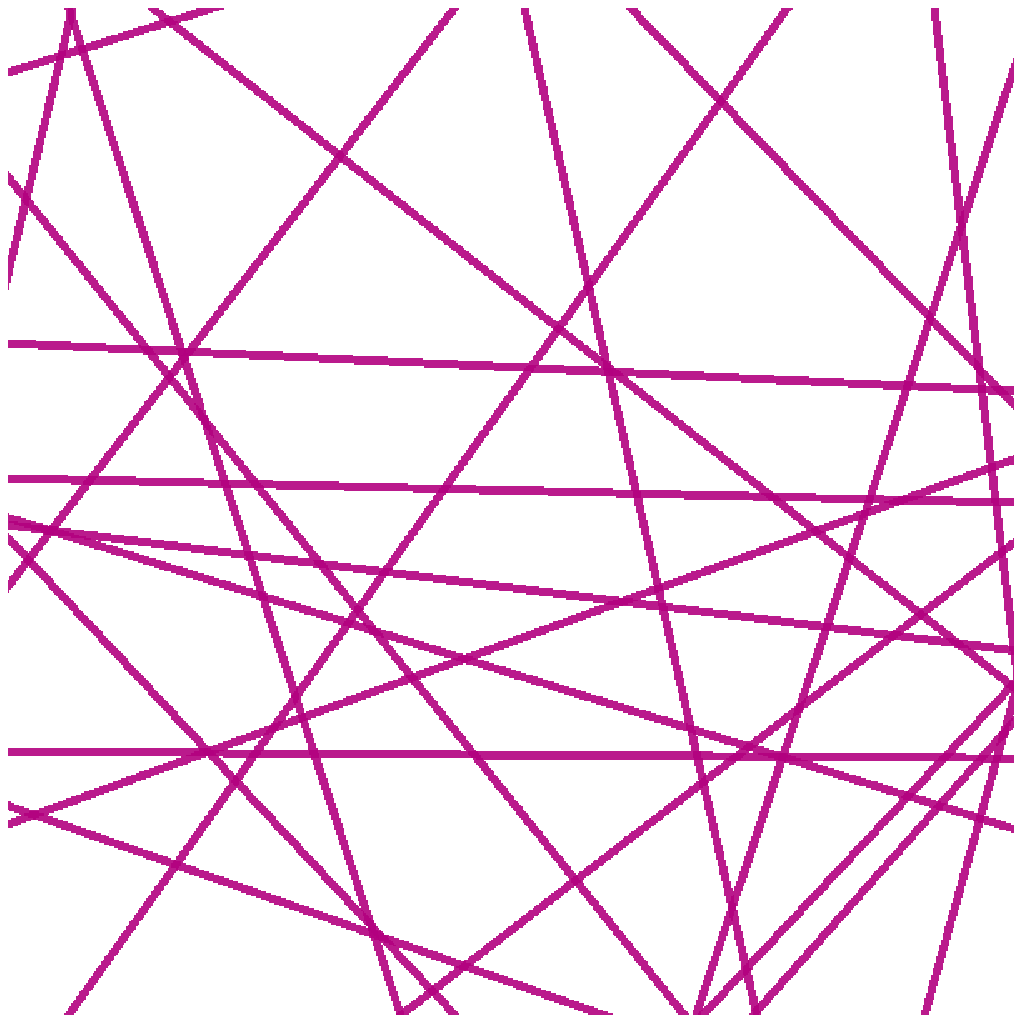}%
%%%\includegraphics[trim = 15mm 80mm 20mm 5mm, clip, width=2.8in]{randomlines}
%{\bf b}~\includegraphics[width=1.25in]{skeletal-muscle-excerpt}
%{\bf b}~\includegraphics[width=1.5in]{muscle_histology}
{\bf B}\includegraphics[width=1.5in,height=1.3in]{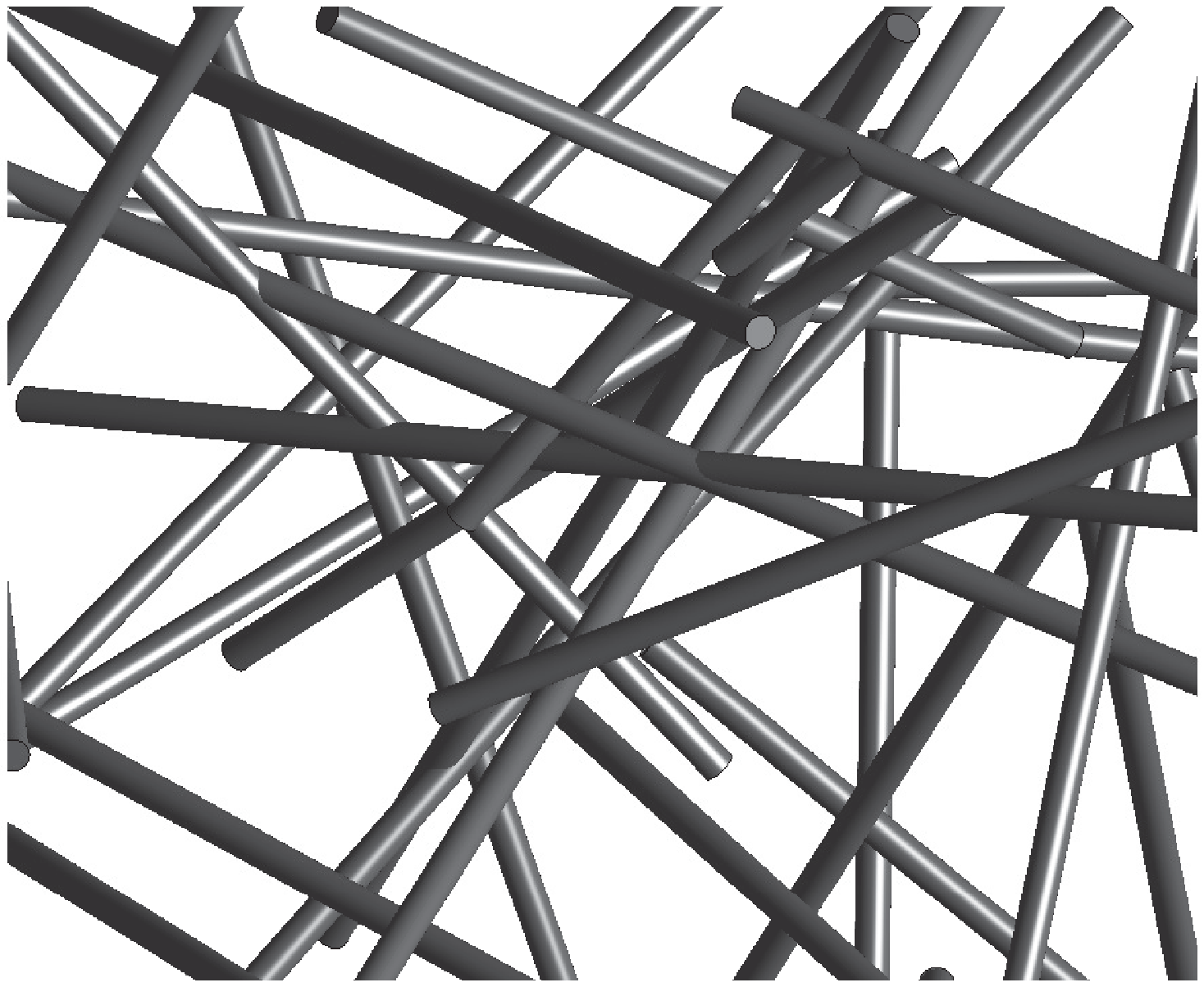}
{\bf C}~\includegraphics[width=1.5in]{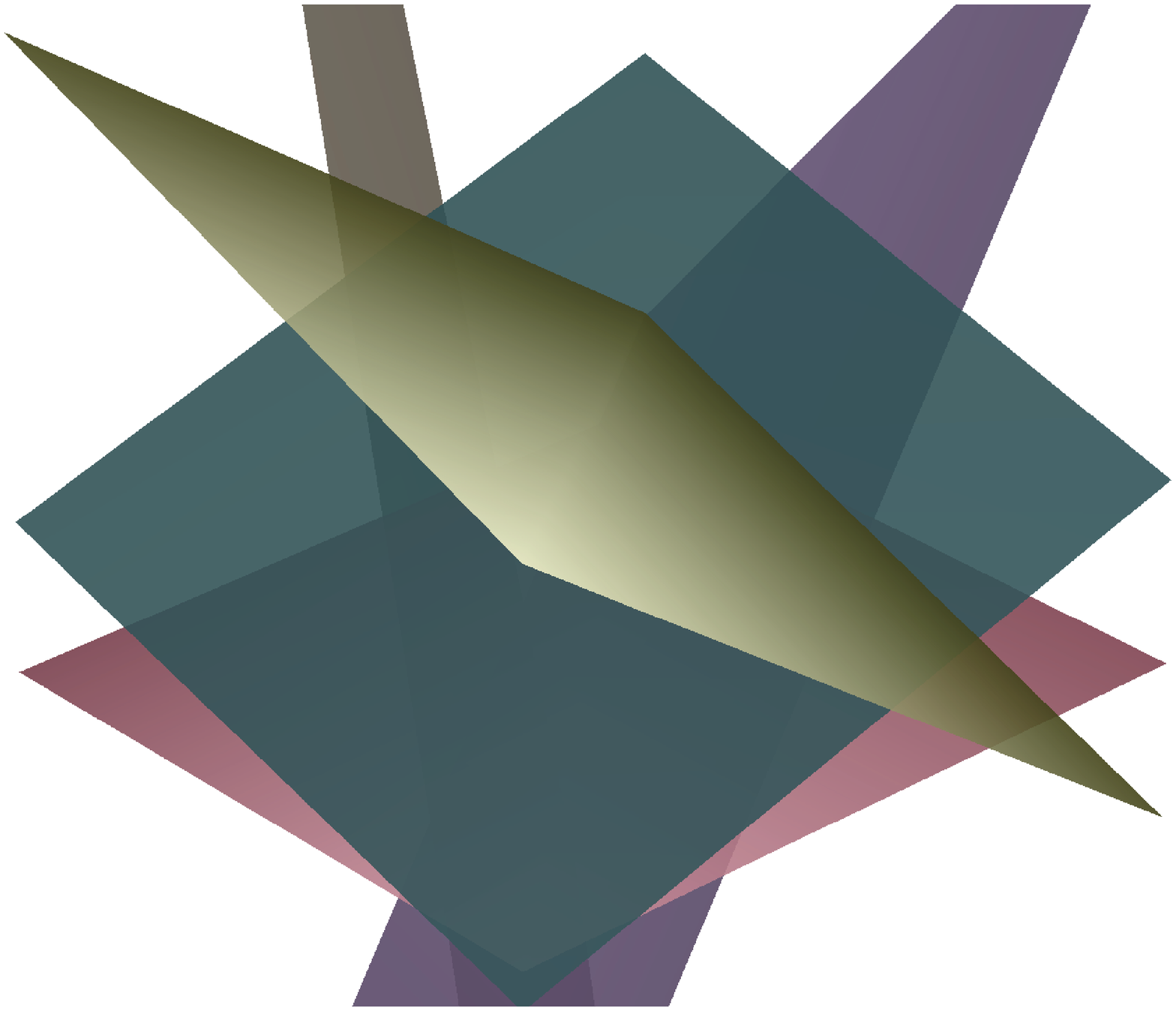}%
{\bf D}~\includegraphics[width=1.7in]{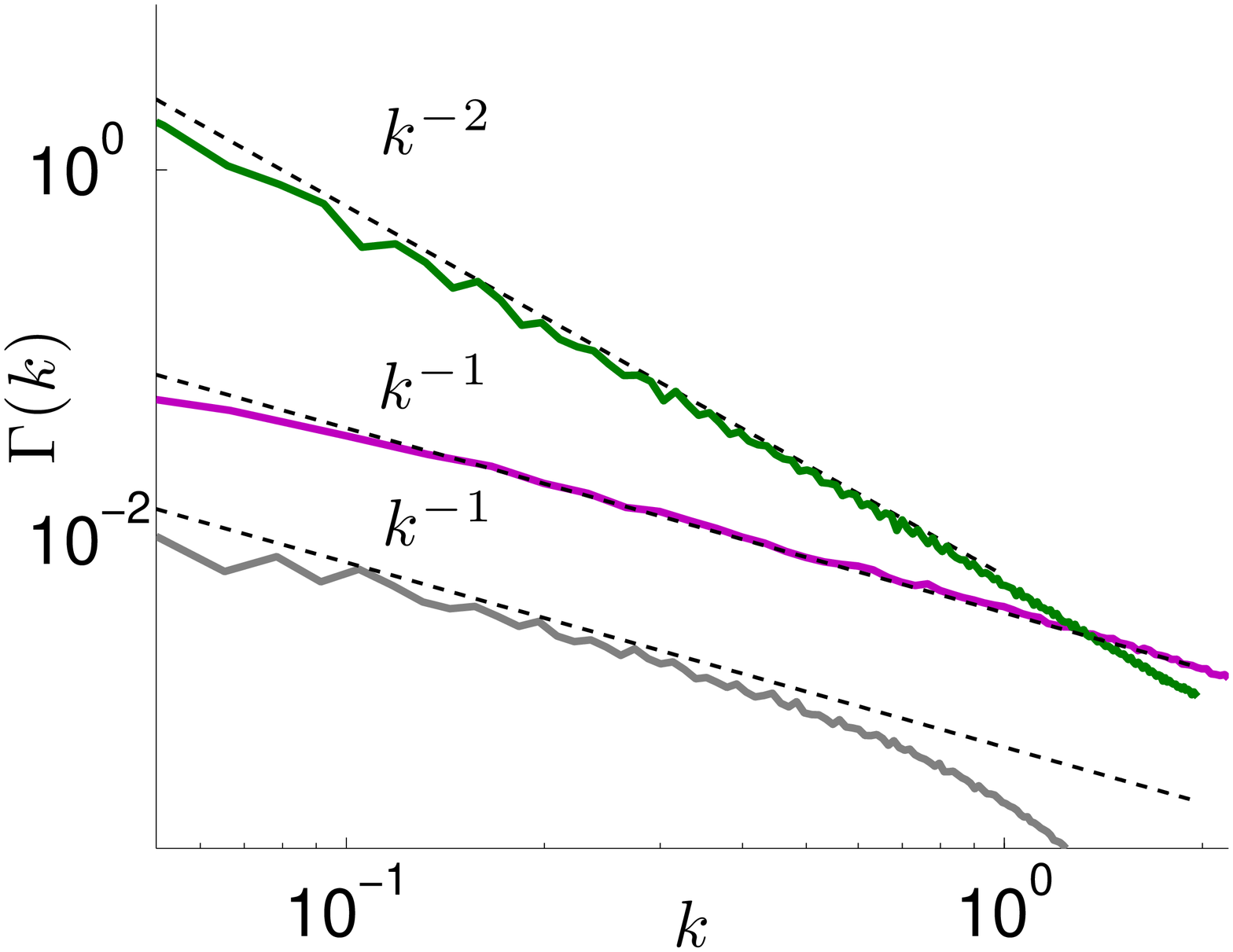}
\caption{Extended structural disorder classes, $d>1$. 
{\bf (A)} Randomly placed and oriented permeable barriers (lines), $d_{s}=1$ in $d=2$ dimensions.
%correspond to the histological slice in {\bf b}, ref.~\onlinecite{Junqueira}, of skeletal muscle across the fibers. Note the tight cell packing achieved by straight cell walls.
{\bf (B)} Randomly placed and oriented rods, $d_{s}=1$ in $d=3$. 
{\bf (C)} Randomly placed and oriented permeable barriers (planes), $d_{s}=2$ in $d=3$.
{\bf (D)} Structure correlator $\Gamma(k)\sim k^{p}$ (numerically calculated and angular averaged, arb. units) for {\bf (A)} (magenta), {\bf (B)} (grey), and {\bf (C)} (green) exhibits the negative exponent $p=-d_{s}$.
}
\label{fig:randomlines}
\end{figure}

Structural complexity (disorder) presents itself in many different forms, e.g. Figs.~\ref{fig:1d} and \ref{fig:randomlines}. 
Our key result is the relation 
\be \label{alpha=p+d}
\alpha= (p + d)/2 
\ee
%\eqref{alpha=p+d} 
(see \cite{supp}) 
between the dynamical exponent $\alpha$ in equations \eqref{vaf} and \eqref{Dinst-main}, and
the structural exponent $p$ characterizing global structural organization in $d$ spatial dimensions. 

The structural exponent $p$ determines the $\Gamma(k)|_{k\to0} \sim k^{p}$ behavior of the Fourier transform of the correlation function $\Gamma(r)$ for the underlying microstructure.
Hence, $p$ characterizes global structural complexity, taking discrete values robust to local perturbations. This enables the classification of the types and topologies of the disorder, 
Figs. \ref{fig:1d} and \ref{fig:randomlines}.

% and for accessing them with diffusion. 
%The classification of transport phenomena based on dynamical exponents has proven instrumental to understanding critical dynamics   \cite{HH} and random Hamiltonians   \cite{Dyson}.
%Our approach extends the use of the dynamical exponent to identifying the restrictions for molecular diffusion.

The relation \eqref{alpha=p+d} provides a way to determine the exponent $p$ and, thereby, the  
structural complexity class,
%the systematic classification of the complexity, in order to study microstructure with 
using any type of bulk diffusion measurement.
%Remarkably, distinct classes of the microstructure reveal themselves 
%in the tail \eqref{vaf} and in the way $D_{\rm inst}(t)$ approaches the finite bulk diffusion coefficient  
%$\Dinf$,
%%%$\Dinf\equiv [\langle \delta x^{2}\rangle/2t]_{t\to\infty}$, 
%equation \eqref{Dinst-main}. 
%Namely, as shown in the Methods section, the dynamical exponent
%in the dynamical exponent
%The relation \eqref{alpha=p+d} ties the diffusion measurement to 
Local properties affect the coefficients, e.g. the values of $\Dinf$ and of the prefactor of $t^{-\alpha}$ in \eqref{Dinst-main}, but not the exponent $\alpha$. The latter  
%determined by the global structural organization, 
is robust with respect to variations between samples of a similar origin.
This picture is akin to critical phenomena   \cite{HH}, where the phase transition temperature is non-universal (sensitive to microscopic details), whereas the critical exponents distinguish, based on global symmetries, between universality classes of long-range fluctuations.
Here, we describe how transport distinguishes between the classes of stationary structural fluctuations.
%and exist for a continuum of disorder densities.

%We apply this framework to water diffusion in tissues.
%Our key observation is that both in muscles (Fig.~\ref{fig:kim} %  \cite{kim-mrm2005} 
%and in the brain (Fig.~\ref{fig:gore}, %  \cite{gore2003}, 
%the dispersion \eqref{vaf}  is unusually slow,  with $\alpha\simeq 0.5$. We interpret this unexpected empirical finding in terms of the universality classes %  \cite{Ernst-I,Visscher,EMT,nphys} 
%of the structural disorder described below from a unified standpoint, which allows us to relate the dispersion \eqref{vaf} to the tissue anatomy at the cellular level.
%The exponent $\alpha$, robust with respect to biological variability, provides a natural way to identify the relevant microanatomy.

%%%%%%%%%%%%%%%%%%%%%%%%%%%%%%%%%%%%%%%%%%%%%%%%%%%%%%%%%%%%%%%%%%%
%%%%%%%%%%%%%%%%%%%%%%%%%%%%%%%%%%%%%%%%%%%%%%%%%%%%%%%%%%%%%%%%%%%

%\subsection*{Qualitative picture}

Figure \ref{fig:1d} illustrates how diffusion distinguishes between the disorder classes via the relation \eqref{alpha=p+d} in $d=1$ dimension. The Monte Carlo simulated diffusion is hindered by the permeable barriers with mean density $\bar n$ and permeability $\kappa$. The structural complexity classes, embodied in the types of barrier placement (Fig.~\ref{fig:1d}a), exhibit qualitatively different structural exponent $p$ in the barrier density correlator (Fig.~\ref{fig:1d}b). 

{\it Order} (periodic placement in Fig.~\ref{fig:1d}a) is reflected in the Bragg peaks in $\Gamma(k)$, with $\Gamma\equiv 0$ for $k$ below the minimal reciprocal lattice vector, formally corresponding to $p=\infty$.  As coarse-graining beyond the largest lattice vector does not contribute to the structural fluctuations, $\D(t)$ decays and $D_{\rm inst}(t)$ reaches $\Dinf$ exponentially fast ($\alpha=\infty$) already at the lattice scale, with the decay rate determined by the lattice specifics.

Structural disorder can be introduced in qualitatively different ways. 
{\it Hyperuniform disorder}  \cite{torquato2003,donev2005} is characterized by suppressed long-range fluctuations,  with the variance in the number of restrictions in a given domain increasing slower than the domain volume  (sub-Poissonian statistics), reflected in $p>0$, and $\alpha>d/2$. Here, we displaced the barriers independently from their positions in a regular lattice by independent random shifts 
%(we took them to be i.i.d. uniformly distributed between $-\bar a/2$ and $\bar a/2$), 
leading to $\Gamma(k)\sim k^{2}$ for small $k$, such that $p=2$ (Fig.~\ref{fig:1d}b), yielding $\alpha=3/2$ (Fig.~\ref{fig:1d}c) according to equation \eqref{alpha=p+d}. 
{\it Short-range disorder} corresponds to the variance of the number of restrictions scaling with the mean number in a given domain, consistent with the central limit theorem. Here, we chose each successive interval $a_{m}$ between barriers independently from the distribution $P(a)$ with mean $\bar a = 1/\bar n$ and finite variance $\sigma^{2}$. This results in the finite plateau $\Gamma|_{k\to0}=\sigma^{2}/\bar a^{3}$, such that $p=0$, qualitatively similar to Poissonian disorder. Hence, $\alpha=1/2$ (Fig.~\ref{fig:1d}c), and, generally, $\alpha=d/2$ in $d$ dimensions  \cite{Ernst-I,Visscher}, cf. equation \eqref{alpha=p+d}.
Finally, {\it strong disorder}, with structural fluctuations growing faster with volume than prescribed by the central limit theorem, is reflected in a diverging $\Gamma|_{k\to 0}$, i.e. the exponent $p<0$, and $\alpha<d/2$ (weak self-averaging). Here, we used the L\'evy (fat tail) distribution $P(a)\sim 1/a^{1+\mu}$ with $\mu=7/4$ for the successive barrier intervals, such that the variance $\langle (a-\bar a)^{2}\rangle_{P}$ diverges. This yields $p=\mu-2=-1/4$ and $\alpha=(\mu-1)/2=3/8$ in agreement with equation \eqref{alpha=p+d}.  
%In the first case, the barriers are placed randomly by choosing each successive interval $a_{m}$ independently from the distribution $p(a)$; we took a lognormal $p(a)$ with mean $\bar a = 1/\bar n$ and variance $\sigma^{2}=(\bar a/2)^{2}$. 
%Finite $\sigma$ corresponds to the finite plateau in the barrier density correlator 
%$\Gamma|_{k\to 0}=\sigma^{2}/\bar a^{3}$, such that the exponent $p=0$ (qualitatively similar to the Poissonian disorder). 

%We emphasize that the qualitative difference between the above disorder types is not immediately obvious from looking at the small fragments 
%(Fig.~\ref{fig:1d}a, blue and green); the hyperuniform fragment may even seem locally more disordered than the short-ranged one. However, the diffusion unequivocally determines that the structural fluctuations in the short-ranged disorder are qualitatively stronger than in the hyperuniform one, leading to a slower decrease of $D_{\rm inst}(t)$, Fig.~\ref{fig:1d}c, confirming the relation \eqref{alpha=p+d}.
%The latter holds even for the ordered barriers (red): as
%In the Supplementary information, we 
%use the exact relation $\Dinf(\bar n) = D_{0}/(1+\bar n D_{0}/\kappa)$ together with equation \eqref{dD-corr} to 
%derive the exact asymptotic limits shown as dashed lines, in agreement with the numerics. 

%%%%%%%%%%%%%%%%%%%%%%%%%%%%%%%%%%%%%%%%%%%%%%%%%%
%%%%%%%%%%%%%%%%%%%%%%%%%%%%%%%%%%%%%%%%%%%%%%%%%%

%The generality of the arguments leading to equation \eqref{alpha=p+d} allows one to {\it classify} the structural complexity (stationary disorder) with diffusion, in terms of the distinct values of the exponent $\alpha$, as
% in terms of the distinct values of the exponent $\alpha$, with the following meaning:
%If the disorder causes the width of the distribution  $\P(D_\nu)|_{L}$, Fig.~\ref{fig:coarsegraining}c,  
%to decrease as
%$\langle (\delta D_\nu)^2\rangle^{1/2} \propto L^{-\alpha}$, the time-dependent diffusion metrics will have the tails  \eqref{vaf} and \eqref{Dinst-main}, correspondingly.
%$D_{\rm inst}(t)-\Dinf\sim t^{-\alpha}$.  
%the diffusion metrics become asymptotically
%sensitive to {global} geometric characteristics. %

Higher dimensions $d>1$ provide more ways to realize the same basic disorder classes. Various  periodic arrangements would yield the same qualitative behavior, $p=\infty$ and $\alpha=\infty$. Hyperuniform disorder can be realized for different $p>0$. While $p=2$ when the restrictions are independently displaced away from the lattice sites,  $p\simeq 1$ for a maximally random jammed state \cite{donev2005}. 
Equation \eqref{alpha=p+d} provides the possibility to observe the jamming transition, from $p=0$ to $p=1$, via diffusion in-between packed impermeable beads.

Remarkably, higher dimensions open up ways to realize strong structural fluctuations, with diverging $\Gamma(k)$, corresponding to $p<0$ and $\alpha<d/2$, without a need to invoke a L\'evy distribution. 
A negative $p$ (Fig.~\ref{fig:randomlines}) can be achieved very naturally, by organizing microstructure in terms of randomly placed and oriented regular components (e.g. infinite lines, planes) with dimensionality $d_{s}<d$, in which case $p=-d_{s}$ (a negative integer), and $2\alpha$ corresponding to their co-dimension. 
The first such example  \cite{nphys} is the extended disorder realized by random permeable hyperplanes, $d_{s}=d-1$, Fig.~\ref{fig:randomlines}a, 
%In this case, $p=1-d$, yielding
%$\langle (\delta D)^2\rangle|_{L} \propto L^{-1}$ similarly to the short-range disorder in $d=1$ dimension, 
resulting in $\alpha=1/2$ in any $d$.
%, SOM \eqref{D-omega-RG}. 
Likewise, randomly placed and oriented rods,  $d_{s}=d-2$, embedded in $d=3$ dimensions (Fig.~\ref{fig:randomlines}c) would realize $p=-1$ and $\alpha=1<3/2$. 
The above examples merely represent each disorder class; e.g. the ``rods'' from Fig.~\ref{fig:randomlines}c can be structurally complex, permeable or impermeable. What matters is the long-range correlations. Clearly, for components with finite extent, the disorder becomes short-ranged, $\alpha\to d/2$,
when the rms molecular displacement exceeds their size.

Above, we assumed that the molecules (the random walkers) can spread everywhere. When impermeable boundaries split the space into disconnected parts, equation \eqref{alpha=p+d} applies separately to the contribution from each part, which then add up. The most relevant disorder contribution is the one with the smallest $\alpha$, yielding the slowest power law tails \eqref{vaf} and \eqref{Dinst-main}. 
%(In view of equation \eqref{Dcum-asy} in the Supplementary Information, the easiest to detect are those with $\alpha<1$.) 

\begin{figure}[t]
\flushleft{\bf A}%
\includegraphics[width=1.7in]{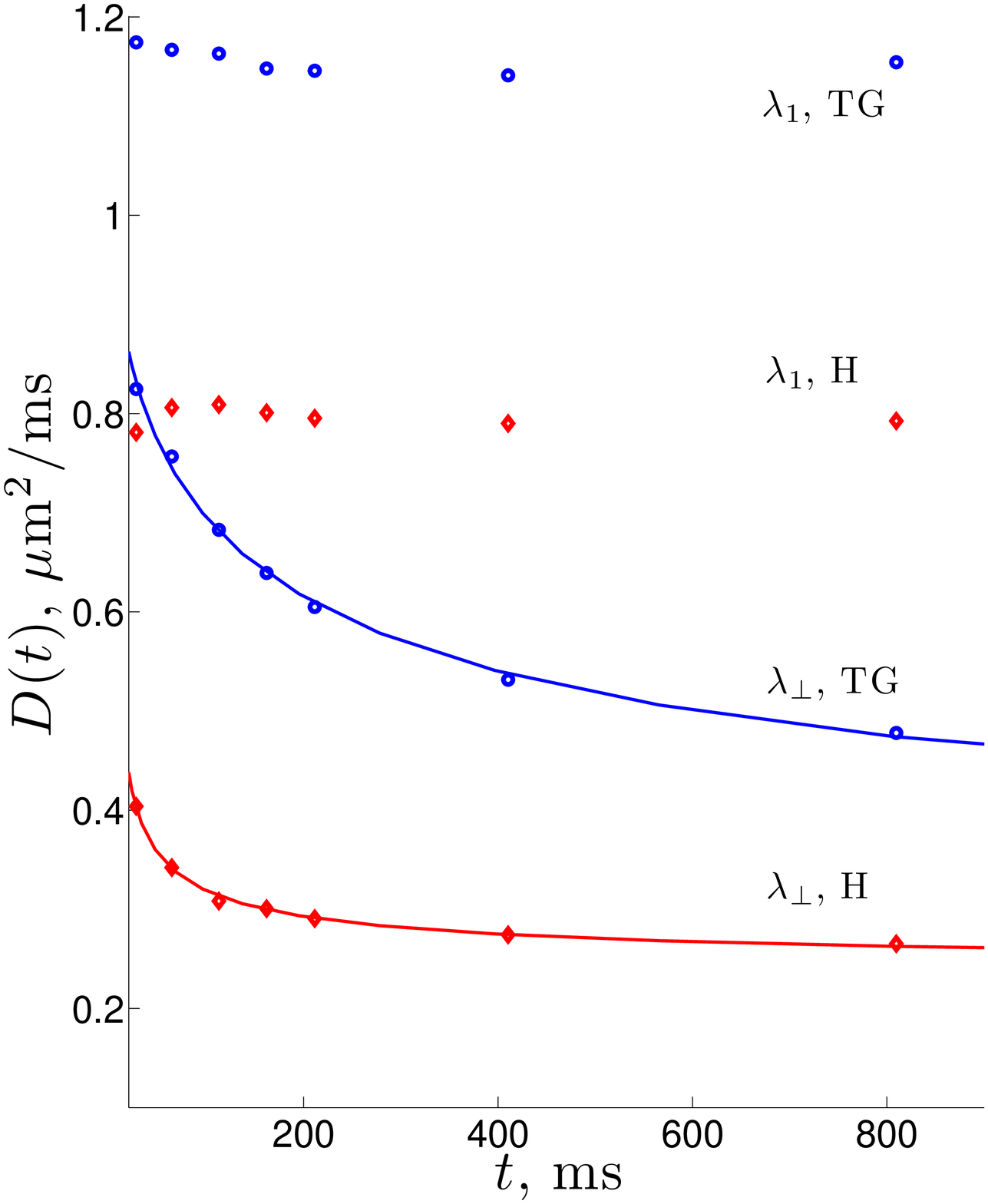}%\qquad
%~~{\bf B}%
%\includegraphics[width=2.2in]{kim-powerlaw_05_1.eps}
{\bf B}%
\includegraphics[width=1.7in]{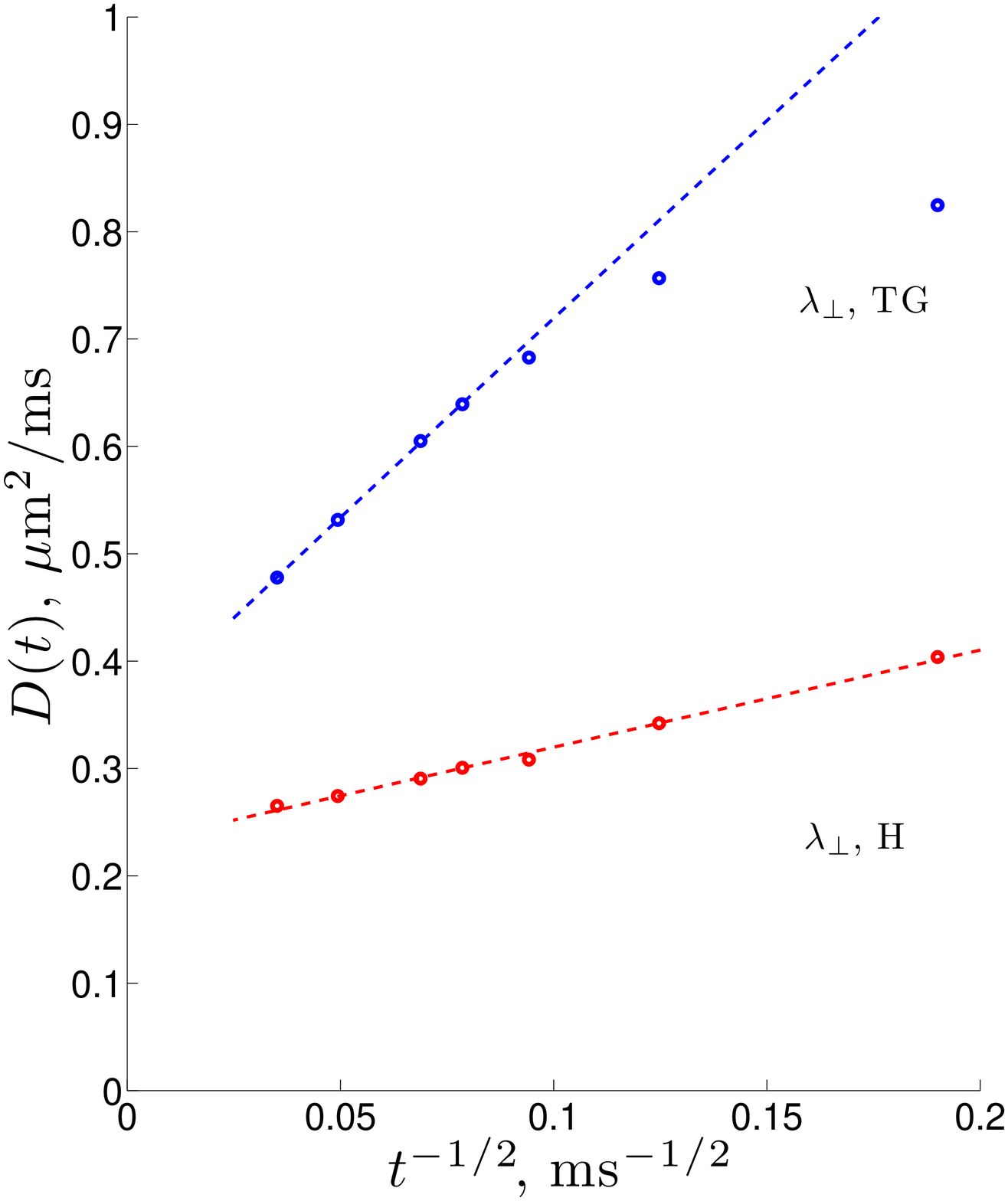}
{\bf C}~\includegraphics[width=1.3in]{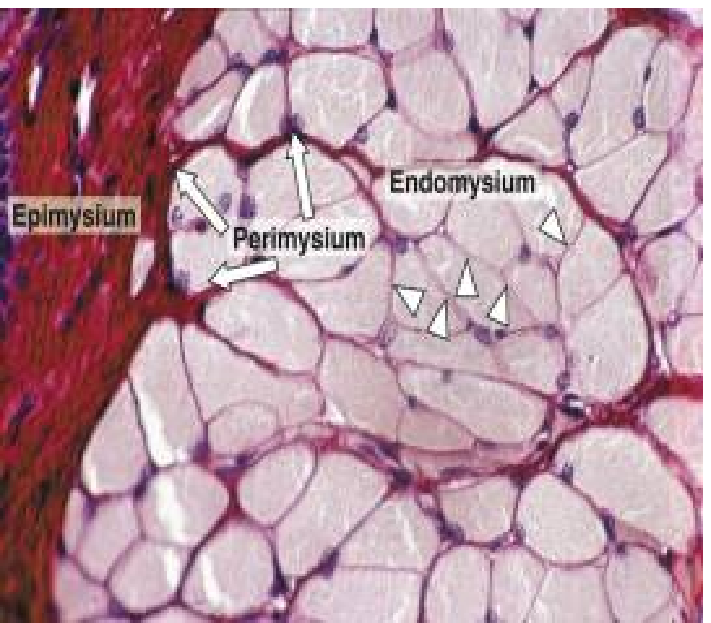}
\caption{
Time-dependent diffusion transverse to muscle fibers  \cite{kim-mrm2005} ($d=2$) reveals
extended structural disorder class of $d_{s}=1$, provided by the muscle fiber membrane (sarcolemma).
{\bf (A)} The longitudinal, $\lambda_{1}$, and the transverse, $\lambda_{\perp}$, diffusion tensor components for calf tongue genioglossus (TG, blue circles) and heart (H, red diamonds).
%The eigenvalues $\lambda_1(t)$ remain nearly constant, while 
%the ones transverse to the fibers ($\lambda_2 \approx \lambda_3$) exhibit notable decrease with time. Here we show 
%the isotropic component $\lambda_\perp = (\lambda_2+\lambda_3)/2$ of 
 %$\lambda_\perp(t)$
%notably decreases with time. 
Solid lines are the fit of $\lambda_\perp(t)$ to Supplementary Eq.~\eqref{D-omega-RG} with $d=2$.
For fit results see Table 1 \cite{supp}.
{\bf (B)} Data for $\lambda_\perp(t)$ replotted as function of $t^{-1/2}$
%. The asymptotically linear dependence on $t^{-1/2}$ for long $t$ is 
consistent with $\alpha=1/2$.  
Equation \eqref{alpha=p+d} yields $p=-1$;  hence, $d_{s}=1$ (see text and Fig.~\ref{fig:randomlines}d).
{\bf (C)} Histological slice  \cite{Junqueira} of skeletal muscle across the fibers. Note the tight cell packing achieved by straight cell walls, qualitatively similar to Fig.~\ref{fig:randomlines}a.
%for $t\gtrsim \tau_{r}$
%(blue and red dashed lines) 
%represents 
%Eq.~\eqref{Dt}, 
%a ``fingerprint'' of cell membranes in the dMRI measurement, cf. Fig.~\ref{fig:randomlines}a,b. 
%The time-dependence is clearly different from $t^{-1}$ (thin black dashed lines) expected from uncorrelated disorder in two dimensions   \cite{Ernst-I}.
%{\it (C)}, 
%Fit of $\lambda_\perp(t)$ to equation \eqref{Dcum-asy}, accounting for experimental errors, 
%for tongue (four longest time points) and for heart (all data points) yields power law exponents $\alpha=0.44\pm 0.30$ and $0.61\pm 0.07$ correspondingly. 
%Large standard error in tongue is due to a relatively narrow range of experimentally available $t$.
%Diffusion in heart, with most permeable membranes, exhibits the dependence \eqref{Dt} practically in the whole measurement range, while for the tongue this time dependence is reached asymptotically.
}
\label{fig:kim}
\end{figure}

\begin{figure}[t]
%\flushleft
\includegraphics[width=3.2in]{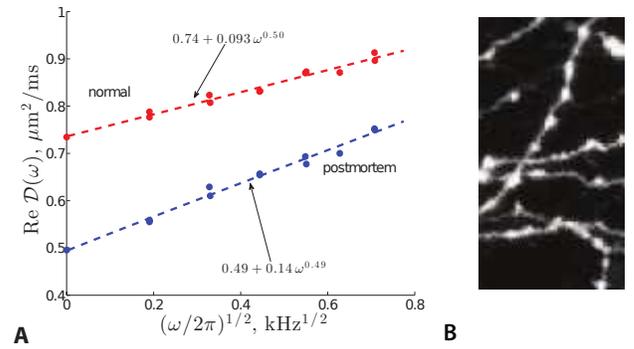}%\quad%
%{\bf b}\includegraphics[width=0.8in]{neurites.eps} 
%{\bf b}%
%\includegraphics[width=2.5in]{does-powerlaw.eps}%\quad
%{\bf c}%
%\includegraphics[width=2.2in]{does-1d.eps}
\caption{Dispersive diffusion in cerebral gray matter %before and after global ischemia 
reveals strong structural correlations. Notably, they remain qualitatively the same after global ischemia.
{\bf (A)} 
Original data   \cite{gore2003} for `d-sin' and `cos' gradient waveforms,
fitted to Supplementary Eq.~\eqref{Dw}, yields
$\alpha=0.50\pm 0.07$ for normal and $\alpha=0.49\pm 0.05$ for postmortem brain,
consistent with the $p=0$ short-range disorder along the one-dimensional neurites (dendrites and axons) in both cases. 
The role of the disorder (the slope) increases after ischemia.
{\bf (B)} Varicose axons from rat hippocampus area CA1, ref.~\onlinecite{shepherd-raastad}, rationalizing the $d=1$ diffusion inside narrow randomly oriented disordered neurites.
%The same as in {\bf a}, 
%The same data and the fits to Supplementary equation \eqref{Dw}, 
%replotted as functions of $\w^{1/2}$.
%{\bf c}, Neurite diffusivity $\D_{1d}(\w)$ from equation \eqref{D1d} and its fits to equation \eqref{D-omega-RG} with $d=1$.
}
\label{fig:gore}
\end{figure}

As a result, measuring the exponent $\alpha$ with any time-dependent diffusion technique allows one to determine the disorder universality class via the exponent $p$ using the relation \eqref{alpha=p+d}.
Let us now apply this framework to diffusion measured with MRI  \cite{callaghan} (dMRI) in tissues. The dMRI in muscles, Fig.~\ref{fig:kim}, reveals strong disorder (weak self-averaging) in $d=2$ with $d_{s}=1$, realized by the sarcolemma; dMRI in the brain, Fig.~\ref{fig:gore}, reveals short-range disorder along the $d=1$ neurites.
% ($\alpha=d/2$).
%we uncover the $\alpha\simeq1/2$ tails in tissues and relate them to the cellular microarchitecture.

%%%%%%%%%%%%%%%%%%%%%%%%%%%%%%%%%%%%%%%%%%%%%%%
%\section*{\large Cell size and membrane permeability in muscles}

%%%%%%%%%%%%%%%%%%%%%%%%%%%%%%%%%%%%%%%%%%%%%%%%%%
%%%%%%%%%%%%%%%%%%%%%%%%%%%%%%%%%%%%%%%%%%%%%%%%%%

In Fig.~\ref{fig:kim}, we analyze the time-dependence of diffusion tensor eigenvalues
in the fresh {\it ex vivo} muscle tissue samples measured by Kim {\it et al.}   \cite{kim-mrm2005}.
The nondispersive eigenvalues $\lambda_1$ correspond to the unrestricted diffusion along the fibers.
The transverse components $\lambda_\perp(t)$ in the two-dimensional fiber cross-section, Fig.~\ref{fig:kim}c,
are  strongly dispersive.
Representing the data as function of $t^{-1/2}$, we observe the asymptotic tail \eqref{Dinst-main}. 
Indeed, 
%in Fig.~2c 
the fit of $\lambda_\perp(t)$ to Supplementary Eq.~\eqref{Dcum-asy}
%$D_\infty + \mbox{const}\cdot t^{-\alpha}$
yields $\alpha\approx 0.5$ for both tongue and heart (Fig.~\ref{fig:kim-supp} \cite{supp}),
%\alpha=0.44\pm 0.30$ and $0.61\pm 0.07$ correspondingly
%The $\alpha$ values are 
exemplifying weak self-averaging, $\alpha<d/2$,
%, contrasted with
%manifestly different from 
in contrast to $\alpha=1$ expected for the $d=2$ short-range disorder.
%For other $d=2$ complexity classes, 
%$\alpha> 1$, even further from the observed values.  
We thus conclude that the restrictions to water diffusion are strongly spatially correlated on the scale of the diffusion length (up to $\sim 30\,\mu$m in this measurement), 
which puts them into the extended disorder class of Fig.~\ref{fig:randomlines}a with $d_{s}=1$.

In \cite{supp}, we argue that the relevant restrictions are in fact muscle cell membranes (sarcolemma), and quantify their permeability and cell size (Table 1 \cite{supp}).
The good agreement between the fit parameters and histological values can be rationalized by comparing a typical histological slice transverse to muscle fibers \cite{Junqueira}
(Fig.~\ref{fig:kim}c) with the random barriers in two dimensions 
(Fig.~\ref{fig:randomlines}a). Tight packing of muscle cells makes the fiber walls fairly flat and spatially correlated even over length scales exceeding typical fiber diameter, qualitatively affecting the 
structural correlations, $\Gamma(k)\sim 1/k$, within the plane transverse to the fibers. 
%evolution of the coarse-grained distribution $\P(D_\nu)|_{L(t)}$.

Hence, the dynamical exponent \eqref{vaf} establishes the effect of cell walls on the MRI signal,  leading to the first non-invasive method to image cell membrane permeability {\it in vivo}, with the potential to correlate with tissue physiology and pathology.
% and be applied in the diagnosis of diseases in soft tissues. 
%Summarizing, the apparent power law tail \eqref{Dcum-asy} with $\alpha\approx 0.5$ in diffusion tranverse to muscle fibers, allows us to establish for the first time the effect of permeable cell membranes on the dMRI signal. We quantify the permeability of cell membranes and the mean cell diameter by using our model   \cite{nphys} which accounts for the singular time dependence \eqref{Dcum-asy}.

%%%%%%%%%%%%%%%%%%%%%%%%%%%%%%%%%%%%%%%%%%%%%%%%
%\section*{\large Diffusivity drop in acute cerebral ischemia}

%%%%%%%%%%%%%%%%%%%%%%%%%%%%%%%%%%%%%%%%%%%%%%%%%%
%%%%%%%%%%%%%%%%%%%%%%%%%%%%%%%%%%%%%%%%%%%%%%%%%%

We now turn our focus to brain, Fig.~\ref{fig:gore}.
The early observation   \cite{moseley'90} of the almost two-fold decrease in the water diffusion coefficient minutes after brain injury has helped to spur the development of {\it in vivo} dMRI. 
Measured at times $t\sim 100\,$ms when any residual time dependence in brain is small, the diffusion coefficient is now widely utilized clinically as a non-invasive diagnostic marker for acute ischemia   \cite{adams'07}.
However, the microscopic origin of this phenomenon has remained under debate for two decades. 
A closely related challenge is to identify the predominant restrictions or cell mechanisms which determine water diffusion in healthy brain.

Here, we address these two related questions by focussing on the Fourier transform $\D(\w)$ of the velocity autocorrelation function \eqref{vaf}
[Supplementary \eqref{Dw}]  in rat cortical gray matter.
We observe that the real part 
%  \cite{EMT,sv-og} 
of  $\D(\w)$  measured by Does {\it et al.} \cite{gore2003} with oscillating gradients   \cite{callaghan} exhibits the $\alpha=1/2$ dispersion (Fig.~\ref{fig:gore}) in the whole frequency range, $\w/2\pi \leq 0.5\,$kHz. 
%Replotting the data   \cite{gore2003} as a function of $\w^{1/2}$ (Fig.~\ref{fig:gore}b) reveals 
%the dispersion \eqref{vaf} with $\alpha=1/2$ to a high accuracy
%for all available $\w$. This is confirmed by the fit to equation \eqref{Dw}
%resulting in $\alpha\simeq 0.5$ both {\it before and after global ischemia} (Fig.~\ref{fig:gore}b).
This value of $\alpha$ is striking for two reasons. First, microstructure must be highly correlated, since,
for the isotropic  \cite{gore2003} diffusion in $d=3$,  the naively expected $\alpha=3/2$
in the absence of structural correlations, $p=0$.  
Second, the value $\alpha=1/2$ is the same {\it before and after global ischemia}.

In \cite{supp}, we radically narrow down the scope of plausible scenarios for the observed change in $\D(\w)$ based on the previously unidentified $\alpha=1/2$ tail \eqref{vaf}, ruling out active streaming breakdown, full confinement by impermeable walls, cell swelling, and increase in the cytoplasmic viscosity as the predominant mechanisms. 
Out of the remaining possibilities, we argue that most of the dispersion arises from the diffusion along randomly oriented narrow neurites (mostly dendrites in gray matter, and, possibly, some axons). Strong apparent correlations are maintained by the impermeable neurite walls such that, effectively, $d=1$, while the disorder along the neurites (such as shown in Fig.~\ref{fig:gore}b) is short-ranged, $p=0$, so that the universality class is that of Fig.~\ref{fig:1d} (blue). 
%From the fit of the neurite component of the signal to Supplementary equation \eqref{D-omega-RG} with $d=1$, 
%the typical distance between those restrictions along the neurites, $a\approx 2-4\,\mu$m, does not change appreciably in the ischemic brain, whereas 

This disorder, for the dendrites, may include   \cite{spines} spines, variations in thickness (``beads''), and in local directionality on the  $\sim 1\,\mu$m scale; for the axons, the synaptic boutons (varicosities) separated   \cite{shepherd-raastad} by $3-6\,\mu$m. 
%The narrow shafts in-between beads serve as regions with reduced local diffusivity, qualitatively behaving as permeable barriers. 
%The dispersion with $\alpha=1/2$ then arises 
% \cite{Ernst-I,Visscher} 
%if the disorder is short-ranged, as we discussed above (Fig.~\ref{fig:1d}. 
Our short-range disorder conclusion is remarkably consistent with the measured variance in the varicosity number within a window growing in proportion to the mean within this window, ref.~\onlinecite{shepherd-raastad}, a defining signature of the $p=0$ exponent. 
Ischemia causes beading, i.e. more pronounced varicosities in both dendrites and axons  \cite{zhang-murphy-2005,li-murphy-2008}, which is likely to increase the disorder.
This is consistent with the increase in the prefactor in the $\w^{1/2}$ contribution to $\D(\w)$,  Fig.~\ref{fig:gore}, as this prefactor generally grows when disorder gets stronger.
Our analysis yields that the effective ``permeability'' of shafts between beads drops more than twofold in ischemia.

Here, our approach underscores the value of the {\it time dependence} of diffusion, rather than of a single number $D_\infty\equiv \D(\w)|_{\w=0}$, for uncovering the origin of a complex biophysical phenomenon. It adds a crucial piece, the short-range disorder, to the picture of impermeable cylinders for the neurites \cite{charmed,Jespersen2007}, and is consistent with the decrease in $\Dinf$ and appearance of beads under a mechanical stress in {\it ex vivo} axons  \cite{budde-frank2010}. The present framework may stimulate more focussed investigations of ischemic stroke, as well as of other neurological disorders. In particular, one could correlate the time-dependent diffusion with the morphological changes during status epilepticus and electrical activation \cite{prichard-1995},  and severe hypoglycemia \cite{hasegawa}, also known to reduce the value of $D_{\infty}$. The reduction in the number of axonal varicosities in Alzheimer's disease relative to the healthy brain  \cite{beads-alz} is likely to result in the decrease of the $\w^{1/2}$ contribution due to the reduced structural disorder.

%%%%%%%%%%%%%%%%%%%%%%%%%%%%%%%%%%%%%%%%%%%%%
%\section*{\large Outlook}

%\nin
To conclude, 
%structural complexity (disorder) presents itself in many different forms. 
%Here, 
we connected the dynamical exponent \eqref{alpha=p+d} to the global structural organization, in order to study microstructure with any type of diffusion measurement.
This framework is particularly useful for biological tissues: While biophysical parameters may vary strongly and continuously between samples, the exponent $\alpha$ takes fixed values determined by the disorder universality class, and is robust with respect to the biological variability.
As a result, we identified the dominant role of cell membranes restricting water motion in muscles, and argued for an increase in the structural disorder along the neurites as a cause of the diffusion coefficient decrease after ischemic stroke. 
%These findings rely on the empirically observed power law dispersion \eqref{Dcum-asy} and \eqref{Dw} with $\alpha\approx 0.5$, signifying the corresponding power law tail in the velocity autocorrelator \eqref{vaf}, which we related to the restricted diffusion with essential spatial correlations in the restrictions. 
We believe the presented classification of the disorder could help identify and quantify the dominant types of restrictions in other living tissues, as well as in classical diffusion or heat or electrical conduction in composite materials, porous media, and other structurally complex samples. Extending this approach to the quantum or wave transport would tie the rich physics of localization with the types of the global structural organization.

%{\bf Acknowledgments.}
It is a pleasure to thank Sungheon Kim, Valerij G. Kiselev and Daniel K. Sodickson for discussions.
Research was supported by the Litwin Fund for Alzheimer's Research, and by the
%National Institutes of Health 
NIH (1R01AG027852).

%%%%%%%%%%%%%%%%%%%%%%%%%%%%%%%%%%%%%%%%%%%%%%%%%%%%%%%%%%%%%%%%%%%%
%\newpage

%\end{article}

%We thank Els Fieremans, Sungheon Kim, Valerij Kiselev and Daniel Sodickson for discussions.
%Research was supported by the Litwin Fund for Alzheimer's Research (JAH), and the
%National Institutes of Health Grant 1R01AG027852 (JAH).
%
% \item[Author Contributions]
%D.S.N. performed calculations and wrote the manuscript.
%J.H.J and J.A.H. supervised the project and edited the manuscript.
%All authors discussed the results and implications and commented
%on the manuscript at all stages.
% \item[Competing Interests] The authors declare that they have no
%competing financial interests.
% \item[Correspondence] Correspondence and requests for materials
%should be addressed to D.S.N.~(email: dima@alum.mit.edu).
%\end{addendum}

%\end{document}

%\newpage

%\newpage

%\section*{}

\newpage

%%%%%%%%%%%%%%%%%%%%%%%%%%%%%%%%%%%%%%%%%%%%%%%%
%%%%%%%%%%%%%%%%%%%%%%%%%%%%%%%%%%%%%%%%%%%%%%%%
\section*{\Large Supplementary Material} 
\appendix

\setcounter{page}{1}

{

%%%%%%%%%%%%%%%%%%%%%%%%%%%%%%%%%%%%%%%%%%%%%%%%
\section{Time-dependent diffusion from structural disorder}

{\bf Diffusion metrics.}
The fundamental quantity, the velocity autocorrelator \eqref{vaf}, is often difficult to measure directly. Instead, there exist a number of equivalent time- or frequency-dependent diffusion metrics, with the relations between them described in detail in refs.~\onlinecite{EMT} and \onlinecite{sv-og}. To interpret various kinds of diffusion measurements, such as dMRI results  \cite{kim-mrm2005,gore2003}, here we outline how the power law tail \eqref{vaf} manifests itself in these metrics. From the outset, we assume the sample to be statistically {\it isotropic}, so that the diffusion metrics are isotropic tensors, and the correlation functions  depend on $r=|\r|$ and $k=|\k|$. 
Generalization to the anisotropic case presents no conceptual difficulty, but makes the presentation more cumbersome.

The instantaneous diffusion coefficient 
%\be \label{Dinst}
%D_{\rm inst}(t) \equiv  {\partial\over\partial t}  {\langle \delta x^2(t) \rangle \over 2} \,,
%\ee
%such that $\D(t) =  \partial_{t} D_{\rm inst}(t)$, 
%approaches the macroscopic diffusion constant $\Dinf$ with the corresponding tail
%\be \label{Dt}
%D_{\rm inst}(t) 
%%\equiv  {\partial \over \partial t} {\langle \delta x^2\rangle \over 2}  
%\simeq D_\infty + \mbox{const} \cdot t^{-\alpha} \,,
%\quad t\to \infty \,. 
%\ee
$D_{\rm inst}(t)$ defined in equation \eqref{Dinst-main} of the main text is the natural metric to study structural correlations, as it quantifies how the spreading of a packet of random walkers is hindered by the microstructure at the time scale $t$. From our perspective, it is a perfect quantity to determine the exponent $\alpha$. However, this is not the most commonly utilized metric in practice.

%$D_\infty\equiv [\langle \delta x^{2}\rangle/2t]_{t\to\infty}$. 
%Here, we generally assume a finite $D_\infty$, i.e. that the diffusion is not anomalous  \cite{Bouchaud}, as is empirically the case in a broad variety of materials and tissues.  
%Our point is that, even with the asymptotically ``trivial'' diffusion at $t\to\infty$, the dynamical exponent $\alpha$ of the transient power-law approach of $D_\infty$ reveals essential information about microstructure.
%  \cite{Ernst-I,Visscher,EMT,nphys}. 

The most commonly reported diffusion coefficient 
\be \label{Dcum}
D(t)\equiv {\langle \delta x^{2}(t)\rangle \over 2t} = \frac1t \int_{0}^{t} D_{\rm inst}(t')\, \d t' 
\ee
describes the dynamics of the {\it cumulative}, rather than instantaneous, mean squared displacement along a particular direction ${\bf \hat x}$ over the diffusion time $t$. 
This is the case both in the dMRI  \cite{callaghan,kim-mrm2005} and in the direct molecular tracking techniques  \cite{kusumi2005}. This definition has a perceived advantage of dividing by time, rather than differentiating with respect to it: clearly, differentiating increases the noise, while dividing does not. 

However, the definition \eqref{Dcum} may mask the exponent $\alpha$.
Indeed, defined in this way, the long $t$ behavior
\be \label{Dcum-asy}
D(t) \simeq \Dinf + \mbox{const}\cdot t^{-\tilde\alpha}\,, \quad \tilde\alpha=\mbox{min}\ \{ \alpha, \ 1\} \,.
%D(t) \simeq \lf  \begin{matrix}     \Dinf + \mbox{const}\cdot t^{-\alpha}\,, & \alpha<1 \,;  \cr  
%						\Dinf + \mbox{const}\cdot t^{-1}\,, & \alpha \geq 1 \,. \end{matrix}  \right.
\ee
%$D(t)$ also has the asymptotic form \eqref{Dt} for $\alpha<1$, whereas
%$D(t) \simeq D_{\infty}+\mbox{const}/t$ for $\alpha\geq 1$.  
In other words, for the tail \eqref{vaf} to be manifest in $D(t)$, it should be sufficiently {\it slow}, $\alpha<1$;  in this case it ``goes through'' the averaging over the increasing interval $t$ in equation \eqref{Dcum}.  
%Hence, the measured $D(t)$ can be directly used to establish the power law dispersion \eqref{vaf}---\eqref{Dt} as long as the power-law tail is sufficiently slow, $\alpha<1$.
In the opposite case, $\alpha\geq 1$, the $t^{-\alpha}$ term in $D_{\rm inst}(t)$ becomes subleading to the $1/t$ term from the integral in equation \eqref{Dcum} converging at short $t$. 

Hence, to practically determine the dynamical exponent $\alpha$, one could first check whether the fit to equation \eqref{Dcum-asy}, using the ``less noisy'' definition \eqref{Dcum}, produces the value $\tilde\alpha<1$. If it does (as in our example of diffusion transverse to muscle fibers), this is it, $\alpha = \tilde\alpha$. In the opposite case, the fit would yield the $1/t$ tail, $\tilde\alpha =1$, which would mask the true value of $\alpha \geq 1$. Then, 
 one must perform the differentiation $D_{\rm inst}(t) = \partial_{t}[tD(t)]$ and obtain $\alpha$ from the fit to equation \eqref{Dinst-main}, with the unfortunate effect of amplifying the measurement noise, as shown by comparing Figs.~\ref{fig:1d} and \ref{fig:1d-all}. 
Practically, this results in more stringent requirements on the signal-to-noise ratio and on the greater number of experimental time points.

There is another useful way of uncovering the exponent $\alpha$, as long as $\alpha<2$,
without the need to take a time derivative. The same power law tail 
\be \label{Dw}
\Re \D(\w) 
\equiv \textstyle{\frac12} \langle v_{-\w} v_{\w} \rangle 
%\equiv \int_0^\infty\! \d t\, e^{i\w t} \D(t)
\simeq D_\infty + \mbox{const} \cdot |\w|^{\alpha} \,, \quad \w\to 0\,,
\ee
persists in the dispersive diffusivity $\D(\w) \equiv \int_0^\infty\! \d t\, e^{i\w t} \D(t)$, 
which is the Fourier transform of the retarded velocity autocorrelator \eqref{vaf}.
The physical meaning of $\D(\w)$ is in relating the current  ${\bf J}_{\w,\r}=-\D(\w)\nabla_{\r}\psi_{\w;\r}$ of the random walkers
to their density gradient, somewhat similar to the dispersive electrical conductivity;  it defines the pole of the diffusion propagator, see refs.~\onlinecite{Ernst-I,Visscher,EMT,nphys,sv-og} and also the discussion below.
Remarkably, there exists a standard dMRI measurement protocol, the oscillating gradient technique \cite{callaghan,gore2003}, which directly measures \cite{sv-og} $\Re \D(\w)$. This is the quantity used in the example of diffusion in cerebral gray matter, Fig.~\ref{fig:gore}.
%Equation \eqref{Dw} ties the tail \eqref{vaf} to the velocity spectral density 
%$\langle v_{-\w}v_{\w}\rangle=2 \Re \D(\w)$ 
%accessible with dMRI using .
%entering the pole of a diffusion propagator  \cite{Ernst-I,Visscher,EMT,nphys,sv-og}.

%%%%%%%%%%%%%%%%%
\begin{figure}[t]
\includegraphics[width=3.4in]{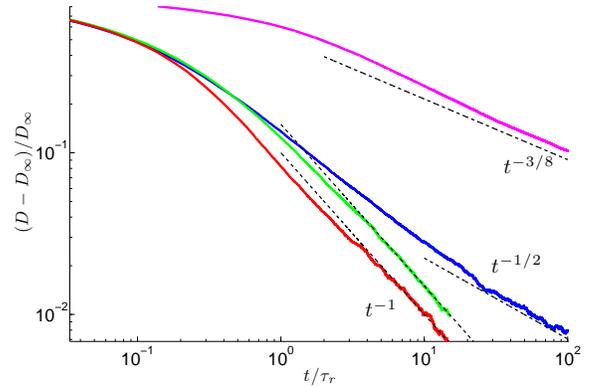}%
\caption{Cumulative diffusion coefficient, equation \eqref{Dcum}, for the one-dimensional example of Fig.~\ref{fig:1d}. Dashed lines correspond to the asymptotic power law decrease of $D(t)$. For $\alpha=1/2$ and $\alpha=3/8$ (short-range disorder, blue, and strong disorder, magenta), the power law in $D(t)$ coincides with that in $D_{\rm inst}(t)$, cf. equation \eqref{Dcum-asy}, whereas for $\alpha>1$ (periodic, red, and hyperuniform, green), it is masked by the $1/t$ term. Taking the derivative 
$D_{\rm inst} = \partial_{t}\big( tD(t) \big)$ reveals the values of $\alpha$ (cf. Fig.~\ref{fig:1d}) but increases the noise.
}
\label{fig:1d-all}
\end{figure}
%%%%%%%%%%%%%%%%%%%%%%%%%%%%%%%%%%%%%%%%%%%%%%%%%%%%%%%%%%%%%%%%%%%

%\section*{Methods}

{
%\small

{\bf Derivation of equation \eqref{alpha=p+d}. Homogenization.}
In this work, we consider the most widespread situation, when a sample has a nonzero macroscopic diffusion constant $D_\infty\equiv [\langle \delta x^{2}\rangle/2t]_{t\to\infty}$, i.e. the diffusion asymptotically becomes normal, or Gaussian. A well-defined macroscopic $\Dinf$, observed in an overwhelmingly broad variety of microscopically heterogeneous samples, attests to the robustness of the diffusion as a Gaussian fixed point with respect to adding the structural complexity (disorder). In this case, a macroscopic sample represents the disorder ensemble, i.e. the system is self-averaging  \cite{aharony-harris}. 
Conversely, the absence of $\Dinf$, e.g. for fractals, near a percolation threshold \cite{Bouchaud,havlin-benavraham}, or for random drifts in one dimension \cite{Sinai,havlin-1989}, signifies so-called anomalous diffusion  \cite{Bouchaud}.

The general relation of the long-time behavior \eqref{vaf} and \eqref{Dinst-main} to the microstructure rests on the homogenization argument: At long diffusion time $t$, the sample, as seen by random walkers traveling over a growing diffusion length $L(t)\equiv \langle \delta x^{2}(t)\rangle^{1/2} \simeq \sqrt{2\Dinf t}$, appears increasingly more uniform due to self-averaging. The sample is being effectively coarse-grained over $L(t)$, such that the strong microscopic heterogeneity is gradually forgotten, and the deviation $\delta D(\r)=D(\r)-\Dinf$ of the smoothly varying coarse-grained diffusion coefficient $D(\r)$ from $\Dinf$ becomes small. This justifies calculating the self-energy part of the disorder-averaged diffusion propagator only to the lowest (second) order \cite{EMT} in the variable component $\delta D(\r)$.
Eventually, the perturbative treatment around $\Dinf$ becomes asymptotically {\it exact}, and the residual deviation 
\be \label{dDw}
{\D(\w)-\Dinf \over\Dinf} \simeq  - {i\w \over \Dinf^{2} d} \int\! {\d^{d}\k\over (2\pi)^{d}}\, 
{\Gamma_{D}(k) \over -i\w + \Dinf k^{2}} \,
\ee
%of the observed $D_{\rm inst}(t)$ from $\Dinf$ 
is given in terms of 
 the Fourier transform $\Gamma_{D}(k) = \int\! \d^{d}\r \, e^{-i\k \r}\, \Gamma_{D}(r)$ of the 
two-point correlation function $\Gamma_{D}(r)=\langle \delta D(\r_{0}+\r) \delta D(\r_{0})\rangle$ 
%of the coarse-grained diffusion coefficient $D(x)$ 
in $d$ spatial dimensions.
Using the relation between $\D(\w)$ and $D_{\rm inst}(t)$,
\be \label{Dinst-Dw}
D_{\rm inst}(t) = \int\! {\d\w\over 2\pi}\, e^{-i\w t}\, {\D(\w)\over -i(\w+i0)}
\ee
(which can be derived using the cumulant expansion as outlined in ref.~\onlinecite{EMT}),
we obtain
\be \label{dD-corr}
{D_{\rm inst}(t) - \Dinf } \simeq {1\over d \Dinf} 
\int\! {\d^{d} \k \over (2\pi)^{d}}\, \Gamma_{D}(k)\, e^{-\Dinf k^{2}t} 
\ee
Equivalently, the latter can be recast in the form 
\be \label{fluct}
D_{\rm inst}(t) \simeq \Dinf + \mbox{const} \cdot \langle (\delta D)^2\rangle|_{L(t)}  \,,
%{D_{\rm inst}(t)-\Dinf\over \Dinf} 
%%\simeq  \frac1d \times
%= \mbox{const} \cdot {\langle (\delta D_\nu)^2\rangle|_{L(t)} \over \Dinf^2} \,.
\ee
where 
$\langle (\delta D)^{2}\rangle|_{L}$
%$\langle \big(\delta D(x_{0})\big)^{2} \rangle_{x_{0}}$ 
is the variance of the Gaussian-smoothed values 
$\delta D(\r)|_{L} = \int\! \d^{d}\r'\,  \delta D(\r+\r')  \, 
e^{-r'^{2}/L^{2}} / (\pi L^{2})^{d/2}$. 
%(As the sample mean of  $ \delta D(x)|_{L}$ is of the order $\langle (\delta D)^{2}\rangle|_{L}$,  it can be neglected when calculating the variance within the adopted accuracy.)
In other words, the diffusion effectively applies a low-pass filter 
$e^{-k^{2}L^{2}/4}$ to the Fourier components of $D(\r)$ and, thus, to its correlator 
$\Gamma_{D}(k)$,
%$\Gamma_{D}(k) = \langle \delta D_{-k}\delta D_{k}\rangle/V$, where $V$ is the sample volume, 
admitting harmonics with progressively smaller wavenumbers $k\lesssim 1/L(t)$. 
As the variance $\langle (\delta D)^{2}\rangle|_{L} \sim L^{-2\alpha}$ decreases due to the smoothing, the measured diffusion coefficient $D_{\rm inst}(t)$ monotonically decreases towards $\Dinf$. 
The power law exponent \eqref{alpha=p+d}
is then directly related to the dimensionality $d$ and the exponent $p$ which determines the $k\to 0$ behavior of $\Gamma_{D}(k)\sim k^{p}$.

We are interested in the spatial correlations 
$\Gamma(r)=\langle n(\r_{0}+\r)n(\r_{0})\rangle$ of the underlying microstructure $n(\r)$ responsible for the heterogeneity of $D(\r)$. Depending on the sample, $n(\r)$ may stand for the  density of grains, traps, barriers and other structural components (e.g. Figs.~\ref{fig:1d} and \ref{fig:randomlines}), which is often strongly heterogeneous at the microscopic scale. Certainly, the coarse-grained $D(\r)$ is not equal to the local average of the strongly varying microscopic diffusion coefficient caused by $n(\r)$. 
However, the statistics of the {\it large scale fluctuations} of $n(\r)$ asymptotically approaches that of the coarse-grained $D(\r)$, such that for $k\to 0$,
 \be \label{Gamma-Gamma}
\Gamma_{D}(k) \simeq C(\bar n)\cdot \Gamma(k) \,, \quad
C(\bar n) = (\partial \Dinf/\partial \bar n)^{2}\,.
\ee
This asymptotically {\it local} relation rests on the self-averaging property which ensures the smooth
% , in the absence of interference effects  \cite{Altshuler-Aronov} in classical diffusion. 
dependence $\Dinf(\bar n)$ on the sample mean ${\bar n}=\langle n(\r) \rangle$ of the restrictions. Hence, after coarse-graining, a {typical small local fluctuation} $\delta D(\r)$ becomes asymptotically proportional to the {typical} small local fluctuation $n(\r)-\bar n$, as long as the self-averaging assumption holds. 
(Conversely, singular dependence $\Dinf(\bar n)$,  e.g. at the percolation threshold, is associated with the lack of self-averaging.) 
In this way, the exponent $p$ characterizes long range correlations in sample's microstructure, $\Gamma(k) \sim k^{p}$, and becomes accessible with a time-dependent diffusion measurement via the relation \eqref{alpha=p+d}. 

Strong self-averaging in $d$ dimensions occurs when 
the variance $\langle (\delta D)^2\rangle|_{L}$ decreases as the inverse ``diffusion volume''
$L^{-d}$, such as for the short-ranged disorder ($p=0$), or faster, as for order or hyperuniform disorder ($p>0$). Weak self-averaging corresponds to the decrease $\sim L^{-2\alpha}$ with a smaller power, 
$0<\alpha<d/2$. For $p\leq -d$, very strong fluctuations destroy self-averaging, a large sample does not represent a disorder ensemble, the macroscopic $\Dinf$ is  undefined, and the present approach fails. Diffusion becomes anomalous \cite{Bouchaud}, with mean squared displacement $\langle \delta x^{2}(t)\rangle \sim t^{2/z}$ as $t\to\infty$ characterized by the dynamical exponent $z\neq 2$, 
see e.g. refs.~\onlinecite{havlin-benavraham,Sinai,havlin-1989}.

%The short-range disorder is characterized by the finite correlation length $l_{c}$, beyond which the disorder correlator $\Gamma(r) \to 0$, and the limit of $\Gamma|_{k\to 0}$ is finite, corresponding to $p=0$. In this case, the variance $\langle (\delta D)^2\rangle|_{L} \propto (l_{c}/L)^{d}$ in equation \eqref{fluct} in the Methods section decreases in the same way as the relative fluctuation of the structural heterogeneity according to the central limit theorem, as long as the diffusion length $L(t)\gg l_{c}$. 
%%The scaling $\langle (\delta D)^2\rangle|_{L}$ with 
%This simple picture of $D_{\rm inst}(t)$ relying on typical, rather than optimal, structural fluctuations (as long as $\Dinf$ is finite) yields $\alpha=d/2$ in $d$ dimensions \cite{Ernst-I,Visscher}. 
%The examples are the first case in Fig.~\ref{fig:1d}, or a medium made of randomly placed domains of two diffusivity values.

}

{\bf One-dimensional disorder universality classes.}
Here we provide the details of the analytical and numerical calculations used to obtain the results in Figs.~\ref{fig:1d} and \ref{fig:1d-all}. 

{\it Monte Carlo dynamics.---}
For each disorder class, Monte Carlo (MC) simulated random walks of  $4\times 10^6$ random walkers evenly split between 40 disorder realizations of $N=1000$ barriers each, were used to average $\langle \delta x^{2}\rangle$ over the paths and over the ensemble.
The total length of each disorder realization could be either smaller or larger than $N\, \bar a$ since barrier intervals were random (as described below). 
The trajectory of each random walker was a sequence of moves in a randomly chosen direction over a distance $\d x = \sqrt{2D_0\d t} = 0.008\, \bar a$ during a time step $\d t$, where $D_{0}$ is the unrestricted (free) diffusion constant. This choice of $\d x \ll \bar a$ ensured that the free diffusion was well simulated within each inter-barrier interval, i.e. the effects of the finite step $\d x$ at the scale of inter-barrier separation were already negligible. In this way, the added barriers can be viewed as the restrictions, or the ``disorder'', for the ideal free one-dimensional diffusion.

The barrier permeability $\kappa$ (the dimensions of velocity) 
determines the dimensionless disorder strength, \cite{nphys} 
\be \label{zeta}
\zeta= {D_{0} \over \kappa\bar a} \,.
\ee
The value $\zeta=1$ was chosen for all disorder realizations and types.
Keeping the same $\bar a$, $D_{0}$ and $\kappa$ for all disorder classes yields the same macroscopic diffusion constant 
$\Dinf=D_{0}/(1+\zeta) = D_{0}/2$ for all simulations.

Transmission across a barrier occurred with probability $P \propto \kappa \d x /D_0 \ll 1$ (ref.~\onlinecite{KM}). 
The total diffusion time was $100\tau_r$, corresponding to a maximum of $1.5625 \times 10^6$ time steps 
per walker, where 
\be \label{tau-r}
\tau_{r}= {V\over \kappa S} \equiv {\bar a \over 2\kappa} 
\ee
is the mean residence time within an average inter-barrier interval characterized by its surface-to-volume ratio $S/V = 2/\bar a$.  
The MC results were carefully calibrated to yield the exact result for $\Dinf$ (with better than 1\% accuracy), as well as using the exact result for $D_{\rm inst}(t)$ for the periodic barriers, cf. equation \eqref{Dper} below and Fig.~\ref{fig:1d}c, red curve.

The random walk simulator was developed in C++. Simulations were carried on the NYU General Cluster, using 120 central processing unit cores simultaneously, within a total time of about 10h per each disorder class.

%Els:
%Monte Carlo dynamics was realized to simulate the one-dimensional diffusion restricted by identical permeable barriers with a fixed mean density $n$ = $1/\mum$ and permeability $\Kappa$ = 1 $\mum/ms$. The barrier spacing (FIG) was either ordered (periodic arrangement), or disordered according to a hyperuniform distribution (randomly displaced over the mean barrier distance around the periodic positions), or according to a lognormal distribution (do you know the specs?). For each arrangement, the diffusion coefficient D(t) was calculated by averaging the displacement variance $\<x^2\>$ over a total of 4 $\times$ 10$^6$ random walkers evenly split between 40 disorder realizations, consisting of each 1001 barriers. The trajectory of each random walker was a sequence of moves in a randomly chosen direction over a distance $dx = \sqrt{2D_0dt}$ during a time step$dt$, with the total diffusion time up to $100\tau_R$, corresponding to a maximum of 1.5625 $\times$ 10$^6$ time steps per walker. Transmission across a membrane occurred with probability $P ? \kappa dx /D_0 ? 1$ (ref Karger paper NMr Biomed). The random walk simulator was developed in C++. Simulations were carried on the NYU General Cluster, using 120 central processing unit cores simultaneously, within a total time of 8h. 

In order to obtain $D_{\rm inst}(t)$ in Fig.~\ref{fig:1d}, the time derivative in equation \eqref{Dinst-main} was calculated using the Savitzky-Golay smoothing procedure written in Matlab, with the 6th order polynomial interpolation over a window increasing with $t$ to suppress the MC noise that becomes relatively more pronounced at longer diffusion times. 

{\it Asymptotic behavior of $D_{\rm inst}(t)$.--- }
Adding the barriers corresponds to the microscopic density of restrictions (disorder) 
$n(x)=\sum_{m=1}^N \delta(x-x_m)$, where $x_{m}$ are the barrier positions.
In the Fourier domain, this results in the density variation $\delta n \equiv n  - \bar n$
\be \label{dnk}
\delta n_{k} = \sum_{m=1}^N e^{-ik x_m} - 2\pi \bar n \delta(k)\,, \quad \bar n = \frac1{\bar a} \,.
\ee
The asymptotic analytical expressions (dashed power law lines in Fig.~\ref{fig:1d}) are based on the relation \eqref{dD-corr} and on the ``locality'' relation \eqref{Gamma-Gamma}
%\be \label{Gamma-Gamma}
%\Gamma_{D}(k) \simeq C(\bar n)\cdot \Gamma(k) \,, \quad
%C(\bar n) = (\partial \Dinf/\partial \bar n)^{2}\,,
%\ee
asymptotically valid for $k\to 0$ in any $d$. 
In our case, 
\be \label{Gamma-n}
\Gamma(k)=\frac1V \, \la \delta n_{-k}\delta n_{k}\ra  
\ee
is the barrier density correlator, and $V$ is the system volume (length).
Using equations \eqref{dD-corr}, \eqref{Gamma-Gamma}, and $\Dinf(\bar n) = D_{0}/(1+\bar n D_{0}/\kappa)$ valid in $d=1$ for any barrier placement, we obtain the exact expression  
for the small relative deviation from $\Dinf$ in one dimension:
\be \label{dD-1d}
{D_{\rm inst}(t) - \Dinf \over \Dinf}  \simeq \lp {\Dinf \over \kappa} \rp^{2} \int\! {\d k\over 2\pi}\, 
\Gamma(k) \, e^{-\Dinf k^{2}t} \,.
\ee
We will now use the general relation \eqref{dD-1d} with $\Gamma(k)$ for the different disorder classes to obtain the corresponding asymptotic behavior, dashed straight lines in Fig.~\ref{fig:1d}c.

{\it Short-range disorder.---}
A simple way to realize short-range disorder is to place barriers sequentially with their successive intervals $a_{m}=x_{m+1}-x_{m}$ i.i.d. (independent identically distributed) random variables chosen from a given probability density function (PDF) $P(a)$ with a finite mean and variance. 
We now relate the limit $\Gamma(k)|_{k\to 0}$, entering equation \eqref{dD-1d}, 
to the parameters of $P(a)$.
The second moment of the barrier density $n_{k}$, entering the correlator \eqref{Gamma-n},  
\[
\sum_{m,m'}^N\la e^{ikx_{m'}} e^{-ik x_m}\ra =
\sum_{m=1}^N\la 1 + \sum_{s=1} e^{-ik (a_1+...+a_s)} + \mbox{c.c.}\ra
\]
with $x_m -x_{m'} = \sum_{j=m'}^{m-1} a_j$, 
is averaged over the disorder
by splitting the double sum, in the limit $N = \bar n V\to \infty$, into three terms, with $m=m'$,
$m>m'$ and $m<m'$, where $s=m-m'$, and using the geometric series formula:
%Subtracting a sharp peak $\la \delta n_{-k}\ra\la\delta n_k\ra$ at $k=0$ with the width vanishing as inverse system size, obtain for the density variance
%$\la\la \rho_{-q}\rho_q \ra\ra = \la\rho_{-q}\rho_q \ra - \la\rho_{-q}\ra\la\rho_q\ra$
\be\non
\la \delta n_{-k} \delta n_k \ra = N \lb 1 + {\tp_{k} \over 1-\tp_{k}} + {\tp_{k}^* \over 1-\tp_{k}^*}
-\frac{2\pi}{\bar a} \delta(k)
\rb
\ee
where $\tp_{k}=\int\! \d a\, e^{-ika}\, P(a)$ is the characteristic function of the PDF $P(a)$, and $^*$ stands for the complex conjugation.
The last term cures the uncertainty at $k\equiv 0$, setting $\Gamma(k)|_{k=0}\equiv 0$, and does not affect the behavior in question at small but finite $k$.
Combining the fractions, we find the barrier density correlator
\be \label{Gamma-tp}
\Gamma(k) = 
 \bar n \cdot  \frac{1-\tp_k \tp_k^*}{(1-\tp_k)(1-\tp_k^*)} \,, \quad k\neq 0\,.
\ee
As the long-time diffusivity behavior depends only on the $k\to 0$ limit of $\Gamma(k)$ (different from the value $\Gamma(k)|_{k=0}=0$), we can represent this limit in terms of the mean and variance of any $P(a)$ via its cumulants,
$\tp_k = e^{-ik\bar a - k^2\sigma^2/2 + ...}$.
%\[
%\tp_q = e^{-iq\bar a - q^2\sigma^2/2 + ...}
%= 1 - iq \bar a - q^2(\bar a^2 + \sigma^2)/2 + ... \,.
%\]
Substituting into equation \eqref{Gamma-tp} and taking the limit $k\to 0$, we obtain the limit
$\Gamma(k)|_{k\to 0} = \sigma^{2}/\bar a^{3}$, as quoted in the main text. Substituting this limit into equation \eqref{dD-1d}, we obtain 
\be \label{1d-poiss}
{D_{\rm inst}(t)-\Dinf \over \Dinf} 
=  \frac1{\sqrt{2\pi}}\,{\sigma^{2} \over \bar a^{2}} \lp {\zeta \over 1+\zeta}\rp^{3/2}
\lp {\tau_{r} \over t}\rp^{1/2} 
\ee
(dashed line on top of the blue MC line in Fig.~\ref{fig:1d}).
The parameter $\sigma^{2}/\bar a^{2}=1/4$ for our choice of the lognormal $P(a)$.
The usage of the lognormal distribution here is not important; any $P(a)$ with finite variance and the same ratio $\sigma/\bar a$ would yield the same asymptotic dependence \eqref{1d-poiss}.

{\it Strong disorder.---}
When the interval PDF has a ``fat tail" so that the variance $\sigma$ diverges, the 
above result \eqref{1d-poiss} does not apply.
%The tail \eqref{p-levy} leads to the power-law frequency dependence of the diffusivity dispersion which depends on $\mu$.
The asymptotic form of the L\'evy distribution tail  \cite{Bouchaud} 
\be \label{p-levy}
P(a) \simeq {C\over a^{1+\mu}} \,, \quad 1 < \mu < 2 \,.
\ee
This choice for $\mu$ ensures the finite mean $\bar a = \int_{0}^{\infty}\! \d a\, a P(a)$ but yields an infinite variance due to the fat tail.  
Practically, we generated the disorder according to the interval PDF \eqref{p-levy} by choosing $a=a_{\rm min}\cdot \xi^{-1/\mu}$, where $\xi$ is the standard random variable uniformly distributed in $[0, \ 1]$, and 
$a_{\rm min} =  \bar a (\mu-1)/\mu$, such that $a>a_{\rm min}$ and $C=\mu a_{\rm min}^{\mu}$.
%For $0<\mu<1$ already the mean diverges, which is too wild, so we do not consider this pathological case; for $\mu>2$ the variance is finite and things are regular.

Plugging the characteristic function  \cite{Bouchaud} for $a>0$,
$\tp_k = e^{-ik \bar a - C'k^\mu [1+i\tan (\pi\mu/2)]}$ into equation \eqref{Gamma-tp}, we find the $k\to 0$
{\it divergency} in the correlator 
\be \label{Gamma-levy}
\Gamma(k)\simeq {2C' \over \bar a^3} \, |k|^{\mu-2} \,, \quad 
C={2C' \over \pi} \Gamma(1+\mu)\sin{\pi\mu \over 2} \,.
\ee
In the second equation above, $\Gamma(1+\mu)$ is Euler's Gamma function. 
Equation \eqref{Gamma-levy} shows that, by using the L\'evy PDF \eqref{p-levy} for the i.i.d. intervals, it is possible to generate a medium with an infrared-divergent density correlator. Of course, this is just one particular way of doing so, but it suffices here, as we need to describe only one member of the corresponding disorder universality class.

The singular barrier correlator \eqref{Gamma-levy}, substituted into equations \eqref{dDw} and \eqref{dD-1d},
causes the power law dispersion with the exponent $\alpha=(\mu-1)/2$,
\bea\label{Dw-levy}
{\D(\w)-\Dinf\over \Dinf} &\simeq&
- \displaystyle{\lp {\zeta\over 1+\zeta}\rp^{\!2}}
{C' (-i\w/\Dinf)^\alpha \over \bar a \cos(\pi\mu/2)}\,, \qquad \\
{D_{\rm inst}(t)-\Dinf \over \Dinf} &\simeq&
\displaystyle{\Gamma(\alpha)\over \pi}
\displaystyle{\lp {\zeta\over 1+\zeta}\rp^{\!2}} {C'\over \bar a (\Dinf t)^\alpha}\,.
%\quad \alpha={\mu-1 \over 2}
\label{Dt-levy}
\eea
Remarkably, the dispersion with the exponent $0<\alpha<1/2$ decreases qualitatively {\it slower} than that for the finite variance. This observation can allow one to determine the power  $\mu=1+2\alpha$ of the L\'evy-stable PDF of intervals from the time-dependent diffusion.
The ``finite-variance" exponent $\alpha=1/2$ is reached for the borderline case of $\mu=2$, separating infinite and finite values of $\sigma$. In this limit, setting $C'\to \sigma^2/2$, equations \eqref{Dw-levy} and \eqref{Dt-levy} correspond exactly to equation \eqref{1d-poiss}.

Finally, we can express the prefactor of $t^{-\alpha}$ in equation \eqref{Dt-levy} solely in terms of the single barrier properties and the exponent $\alpha$, by excluding the L\'evy tail normalization $C$:
\be \label{Dt-levy-alt}
{D_{\rm inst}(t)-\Dinf \over \Dinf} 
\simeq
{\Gamma(\frac12 -\alpha) \over \sqrt{2\pi}\, \alpha}
\lp {\alpha \sqrt{2} \over 2\alpha+1}\rp^{\!\! 2\alpha+1} \!\!\!\!
\lp {\zeta\over 1+\zeta}\rp^{\!\! 2-\alpha} \!\!\!\!
\lp {\tau_{r} \over t}\rp^{\!\!\alpha}\!\!.
\ee
Equation \eqref{Dt-levy-alt} for $\alpha=3/8$ agrees well with the MC simulations in Fig.~\ref{fig:1d}c.
The cumulative $D(t)-\Dinf$, obtained from this equation using the relation \eqref{Dcum} [which amounts to dividing by $1-\alpha$], agrees well with the corresponding cumulative MC-generated diffusion coefficient in Fig.~\ref{fig:1d-all}.

{\it Hyperuniform disorder.---}
We realize this disorder class in $d=1$ by displacing the barriers from their positions in a periodic arrangement. Now it is the {\it displacements} that are i.i.d. random variables taken from the PDF $P_{\rm displ}(\xi)$. Clearly, the long-range order is preserved, as the system on average ``remembers'' about its ideal lattice positions. Hence, one expects the structural fluctuations to be qualitatively less pronounced as compared to the disorder types described above.

To calculate the correlator \eqref{Gamma-n} for the hyperuniform placement, we begin from the density \eqref{dnk}, where now $x_{m} = m\bar a + \xi_{m}$, and $\xi_{m}$ are the i.i.d. displacements generated according to $P_{\rm displ}(\xi)$. Using $\delta(k)|_{k=0} = V/2\pi$ where $V$ is the system length and 
$\sum_{m} 1 \equiv N = \bar n V$, we obtain
\bea\non
\delta n_{-k}\delta n_{k}  &=& \sum_{m,m'=1}^{N} e^{ik(m-m')\bar a + ik(\xi_{m}-\xi_{m'})} 
\\ \label{nn-hu}
& & + (2\pi\bar n)^{2} {V\over 2\pi} \delta(k) - 2\cdot 2\pi\bar n N \delta(k) \,. \qquad
\eea
Averaging of the double sum over the disorder is done by splitting it into 
the part with $m=m'$ yielding $N$,
and the part with $m\neq m'$
yielding the Debye-Waller factor $\la e^{ik(\xi_{m}-\xi_{m'})}\ra \equiv |\tp_{{\rm displ}, k}|^{2}$,
where $\tp_{{\rm displ}, k} = \int\! \d \xi \, e^{-ik \xi} P_{\rm displ}(\xi)$, multiplied by
\bea \non
\sum_{m, m'=1}^{N} e^{ik\bar a (m-m')} - N &=& \lb {2\pi\over \bar a} \sum_{m} \delta\lp k-k_{m}\rp\rb^{2} - N
\\ \non
&=& N\lb {2\pi \over \bar a} \sum_{m}  \delta\lp k-k_{m}\rp -  1\rb.
\eea   
The sums in the right-hand side span over all the reciprocal lattice vectors $k_{m}={2\pi m/\bar a}$, 
$m=0,\pm 1, \pm 2, ...$, 
which is a consequence of the Poisson summation formula 
$\sum_{m}e^{imk\bar a } = {2\pi \over \bar a} \sum_{m} \delta\lp k - k_{m}\rp$ valid in the limit $N\to \infty$.
Putting all the pieces together, we obtain
\be \label{Gamma-hu}
\Gamma_{\rm hu}(k) = \bar n \lb 
1 - |\tp_{{\rm displ}, k}|^{2}  + {2\pi\over \bar a} \sum_{m\neq 0} 
\delta\lp k -  k_{m} \rp \left| \tp_{{\rm displ}, k_{m}}\right|^{2} 
\rb\!.
\ee
This correlator is familiar from the X-ray scattering in crystals: a series of spikes with decreasing amplitude, together with the ``incoherent'' background determined by the displacement PDF. 
Using the cumulant form of $\tp_{{\rm displ},k} = e^{-\sigma_{\rm displ}^{2} k^{2}/2 + \dots}$, 
we obtain the universal $k\to 0$ behavior 
\be \label{Gamma-hu-asy}
\Gamma_{\rm hu}(k)\simeq \bar n \sigma_{\rm displ}^{2} k^{2} \,. 
\ee
Substituting \eqref{Gamma-hu-asy} into equation \eqref{dD-1d}, we finally obtain
\be \label{1d-hu}
{D_{\rm inst}(t)-\Dinf \over \Dinf} 
=  \frac1{\sqrt{2\pi}}\,{\sigma_{\rm displ}^{2} \over \bar a^{2}} \lp {\zeta \over 1+\zeta}\rp^{1/2}
\lp {\tau_{r} \over t}\rp^{3/2} .
\ee
In our MC simulations, we took
$P_{\rm displ}=1/\bar a$ for $-\bar a/2 <\xi < \bar a/2$ and zero otherwise; substituting its variance 
$\sigma_{\rm displ}^{2} = \bar a^{2}/12$ into equation \eqref{1d-hu}, we obtain the dashed line agreeing with the green MC curve in Fig.~\ref{fig:1d}.

{\it Order (periodic lattice).---}
The ordered (periodic) case admits the exact solution  \cite{sukstanskii-jmr2004,dudko} for the diffusion propagator $G_{\w,q}$ in terms of the Bloch waves. In our present notation, with $a=\bar a$ the lattice period and $\zeta=2\ell/a = D_{0}/\kappa a$, it reads
\be \label{G-per}
G_{\w,q} = \sum_{n=0}^{\infty} 
\frac{2\zeta q^{2} k_{n}^{2} }{\lp 1+\frac2\zeta\rp + k_{n}a \cot k_{n}a}
\frac1{\lp k_{n}^{2} - q^{2}\rp^{2}} \frac1{-i\w + D_0 k_{n}^{2}} 
\ee
where $k_{n}$ are the positive roots of the equation $\cos k_{n}a = \cos qa + k_{n}\ell \sin k_{n}a$.
One readily checks that the $n=0$ term of the sum \eqref{G-per} yields the macroscopic propagator $1/(-i\w + \Dinf q^{2})$ as $q\to0$. 
The contribution of the even terms, $n=2,4,6,\dots$, is $\O(q^{4})$ and does not affect the diffusion coefficient. We now focus on the $\O(q^{2})$ contribution from the odd $n$ terms:
\bea \non
G_{\w,q} &\equiv& \frac1{-i\w + \D(\w)q^{2} + \O(q^{4})} \\ 
&= & \frac1{-i\w + \Dinf q^{2}} + q^{2} F(\w) + \O(q^{4})
\non
\eea
yielding $\D(\w) = \Dinf + \w^{2}F(\w)$, where 
\[
F(\w) = \sum_{n=1,3,...} 
\frac{2\zeta}{k_{n}^{2}} \frac1{\lp 1+\frac2\zeta\rp + k_{n}a \cot k_{n}a}
 \frac1{-i\w + D_0 k_{n}^{2}} 
\]
and $\tan ({k_{n}a}/2) = - \zeta k_{n}a/2$ since $q\to 0$.
Using equation \eqref{Dinst-Dw}, we obtain $D_{\rm inst}(t) = \Dinf - \partial_{t}F(t)$, such that
\be \label{Dper}
D_{\rm inst}(t) = \Dinf + D_{0}\!\!
\sum_{n=1,3,...} \!\!{2\zeta^{2}  \, \exp(-t/t_{n})\over 1+\zeta+(\zeta k_{n}a/2)^{2}} , 
\quad \frac1{t_{n}} = D_{0}k_{n}^{2} .
\ee
This expression is exact for all $t$. For long $t\gtrsim t_{D}=a^{2}/2D_{0}$, where $t_{D}$ is the time to diffuse across one interval, already the first term suffices (the others decay exponentially faster), agreeing very well with the numerical result in Fig.~\ref{fig:1d}c.

We note that the {\it residence time} in one box, $\tau_{r} = \zeta t_{D}$, does not enter equation \eqref{Dper}, so that the times $t_{n}$ are instead determined by $t_{D}$.  
This is remarkable, as it is $\tau_{r}$ that is the time scale that determines the physics of transport at large temporal and spatial scales for any disordered case considered above (and $\tau_{r}\gg t_{D}$ for weakly permeable barriers, $\zeta\gg 1$). This observation suggests that {\it perfectly ordered systems are exceptional}, in the sense that the time dependence of transport in them is not representative of most biological and man-made samples that are at least somewhat disordered. Diffusion (and transport in general) in perfectly ordered samples exhibits coherence due to the infinitely long spatial correlations. Physically, for the system to equilibrate (establish a long-term density profile) it is enough for the density of random walkers to equilibrate within each identical interval; there is no need to hop over barriers to sense the density in neighboring intervals. 

\begin{figure}[t]
\flushleft{\bf a}%
\includegraphics[width=1.7in]{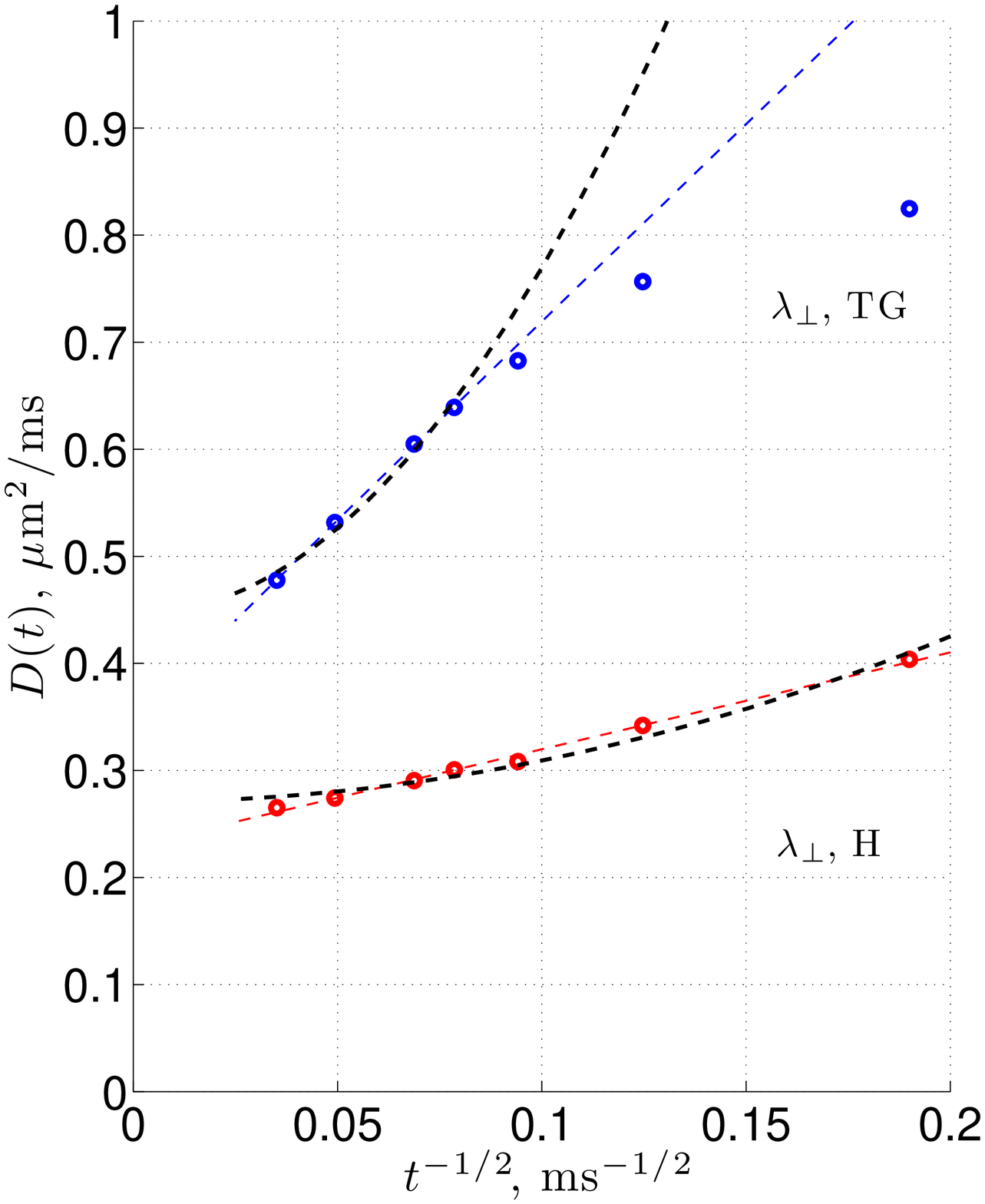}%
{\bf b}%
\includegraphics[width=1.7in]{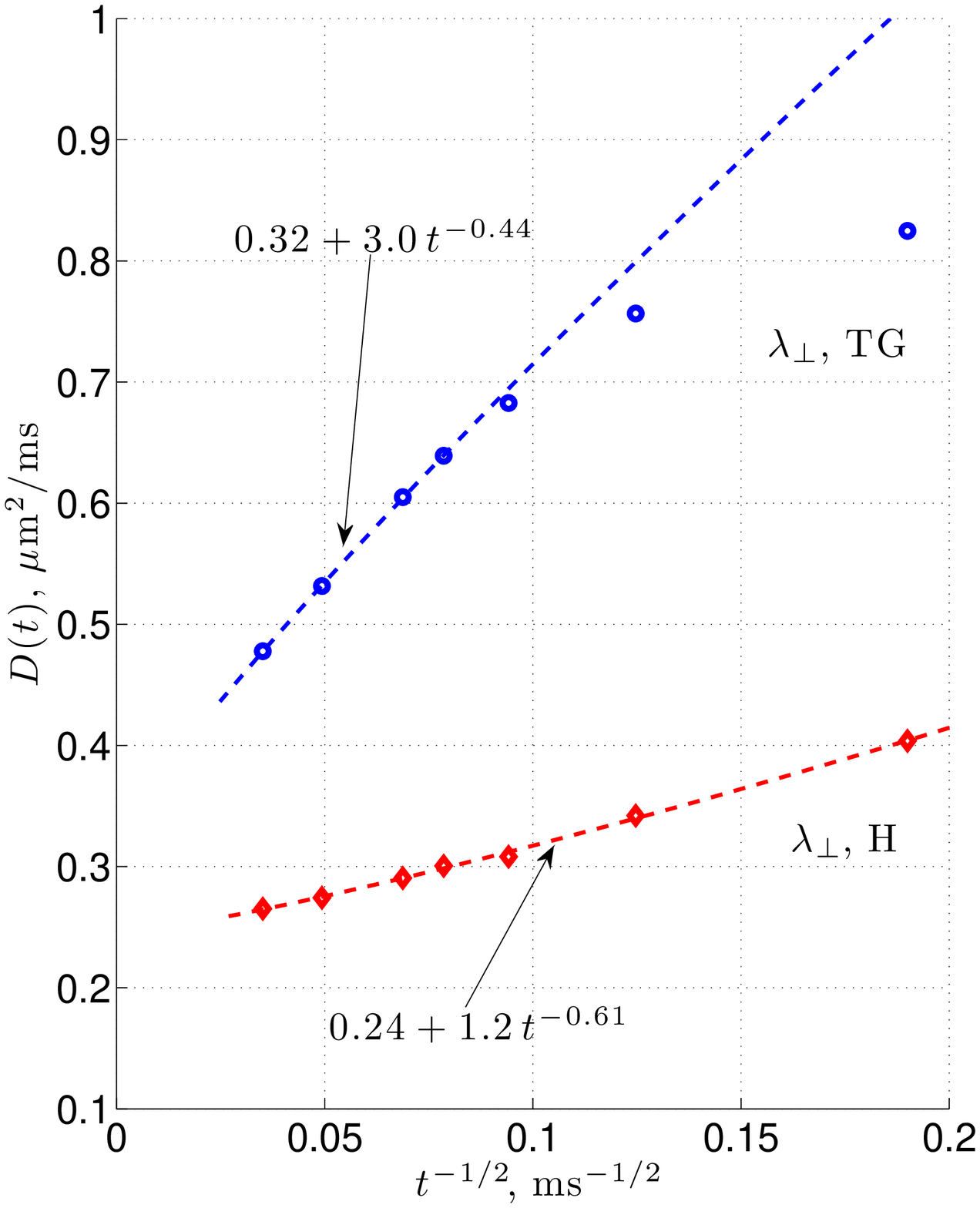}%
\caption{The asymptotic power-law tail in the time-dependent diffusion in muscle fibers from ref.~\onlinecite{kim-mrm2005}.
{\bf (A)}, 
Comparing the $t^{-1/2}$ (thin blue and red dashed lines) and $t^{-1}$ (thick black dashed lines) power laws for calf tongue genioglossus  (TG) and heart (H), cf. Fig.~\ref{fig:kim}b.
%The eigenvalues $\lambda_1(t)$ remain nearly constant, while 
%the ones transverse to the fibers ($\lambda_2 \approx \lambda_3$) exhibit notable decrease with time. Here we show 
%the isotropic component $\lambda_\perp = (\lambda_2+\lambda_3)/2$ of 
 %$\lambda_\perp(t)$
%notably decreases with time. 
{\bf (B)}, 
Three-parameter fit of $\lambda_\perp(t)$ to the power law  \eqref{Dcum-asy}, accounting for experimental error bars, 
for tongue (four longest time points) and for heart (all data points) yields power law exponents $\alpha=0.44\pm 0.30$ and $0.61\pm 0.07$ correspondingly. 
Large standard error in tongue is due to a relatively narrow range of experimentally available $t$.
%Diffusion in heart, with most permeable membranes, exhibits the dependence \eqref{Dt} practically in the whole measurement range, while for the tongue this time dependence is reached asymptotically.
}
\label{fig:kim-supp}
\end{figure}
%%%%%%%%%%%%%%%%%%%%%%%%%%%%%%%%%%%%%%%%%%%%%%%%%%

%%%%%%%%%%%%%%%%%%%%%%%%%%%%%%%%%%%%%%%%%%%%%%
\begin{table*}[t]
\caption{Fit results for transverse eigenvalues $\lambda_\perp(t)$ in heart (H) and tongue genioglossus (TG) muscles  \cite{kim-mrm2005}, Fig.~3.
First three columns are the fit parameters using $D(t)$ obtained from equation \eqref{D-omega-RG}, the rest are the quantities derived from them. }
%For the heart, the unrestricted diffusion coefficient $D_0$ is fixed to the corresponding non-dispersive longitudinal eigenvalue $\lambda_1$.
%For TG, the fit parameter $D_0$ agrees well with the non-dispersive eigenvalue $\lambda_1\approx 1.16\,\mu$m$^2$/ms, see Fig.~2.
\begin{tabular}{|c|c|c|c||c|c|c|c|c|c|c|}
%\begin{ruledtabular}
\hline
        &  $D_0$, $\mu$m$^2$/ms  & $\zeta$   &  $\tau$, ms &  $D_\infty$, $\mu$m$^2$/ms  &  $\tau_D$, ms &  $\tau_r$, ms   &  $\ell$, $\mu$m  &  $S/V$, $\mu$m$^{-1}$   & $\kappa\times 10^2$, $\mu$m/ms  & $a$, $\mu$m \\
\hline
\hline
H     &0.8 (fixed)&2.59   &68.8  &0.223  &20.5    &26.5   &7.42  &0.699   &5.39    &5.72
\\
TG     &1.09       &2.84   &1328  &0.283  &329     &467    &38.0  &0.150   &1.43    &26.7
\\
\hline
\end{tabular}
%\end{ruledtabular}
\end{table*}
%%%%%%%%%%%%%%%%%%%%%%%%%%%%%%%%%%%%%%%%%%%%%%

%%%%%%%%%%%%%%%%%%%%%%%%%%%%%%%%%%%%%%%%%%%%%%%%
\section{Permeability and cell size in muscle fibers}

%The strong time dependence of in regions of calf tongue and heart for diffusion times $30\,\mbox{ms}<t<800\,$ms is due to complexity of tissue microarchitecture.
Here we focus on the eigenvalues $\lambda_{2}(t)$ and $\lambda_3(t)$ [notation of ref.~\onlinecite{kim-mrm2005}] transverse to muscle fibers in the heart (H) and in the tongue genioglossus (TG).
(We do not consider here the case of tongue core also measured in ref.~\onlinecite{kim-mrm2005}, where the geometry is more complicated than that of parallel fibers.) 

As diffusion is axially symmetric  \cite{kim-mrm2005},
$\lambda_2(t) \approx \lambda_3(t)$ both in H and TG,  we consider the isotropic transverse component
$\lambda_\perp=(\lambda_2 + \lambda_3)/2$, contrasted in Fig.~\ref{fig:kim}a with the practically nondispersive eigenvalues $\lambda_1$ along the fibers.

Fig.~\ref{fig:kim-supp} demonstrates that the asymptotic $t\to\infty$ behavior of $\lambda_{\perp}(t)$ is consistent with equation \eqref{Dcum} with $\alpha=1/2$. 
(The analysis of $\lambda_2(t)$ and $\lambda_3(t)$ separately, not shown here, yields similar results.)
In particular, Fig.~\ref{fig:kim-supp}a demonstrates that the competing possibilities of $\alpha\geq 1$ are not consistent with the data; $\alpha=1$ would correspond to 
$\lambda_{\perp}(t)\simeq D_{\infty}+ \mbox{const} \cdot (\ln t)/t$ according to equation \eqref{Dcum}, whereas $\alpha>1$ would result in 
$\lambda_{\perp}(t)\simeq D_{\infty}+ \mbox{const}/t$ according to equation \eqref{Dcum-asy}.

In this way, the dynamical exponent \eqref{alpha=p+d} helps identify the $d_{s}=1$ restrictions to the $d=2$ dimensional diffusion. To quantify the physical parameters of the underlying muscle fiber membrane, one needs a model for the time dependent diffusion beyond the asymptotic long-time regime. 
Recently, we found the corresponding dispersive diffusivity 
for the random barriers (hyperplanes) in any dimension $d$: 
%relevant for quantifying the dMRI measurements \cite{kim-mrm2005,gore2003} in tissues,
%we apply the expression
%Recently, we established that the divergence $\Gamma \sim |k|^{1-d}$ for $d>1$ can be realized in a much more conventional way, in a class of extended structural disorder  \cite{nphys}, with
%a simplest representative being a set of infinite flat membranes ($d-1$-dimensional planes) with permeability $\kappa$ and a given $S/V$ ratio, randomly placed and oriented in $d$ dimensions. The membranes divide a sample into ``cells'' of random shapes (see Fig.~\ref{fig:randomlines}a for $d=2$). 
%A key signature of the diffusivity $\D(\w)$, 
\be \label{D-omega-RG}
{D_0 \over \D(\w)} = 1 + \zeta + 2z_\w(1-z_\w)\lb \sqrt{1 + \zeta/(1-z_\w)^2} - 1\rb 
\ee
%for the dispersive diffusivity $\D(\w)$,
%found in 
ref.~\onlinecite{nphys}, approximately for all $\w$, permeability $\kappa$ and surface-to-volume ratio $S/V$. 
Its low-frequency behavior is indeed characterized by the exponent $\alpha=1/2$ 
in any dimensionality $d$. 
Here, %as in  ref.~\onlinecite{nphys}, 
$D_0$ is the unrestricted diffusion coefficient, 
$\zeta=S\ell/Vd$ [cf. equation \eqref{zeta}], 
$2\ell = D_0/\kappa$, $z_\w = i\sqrt{i\w\tau}$, and $\tau = \ell^2/D_0 = D_0/(2\kappa)^2$. 
The corresponding expression for $D(t)$ is obtained using numerical integration procedure described in ref.~\onlinecite{nphys}.

This allows us to move one step further and quantify the membrane permeability and surface-to-volume ratio. 
The fit of the in-plane diffusivity $\lambda_\perp(t)$ to the time dependent diffusion coefficient $D(t)$ obtained from equation \eqref{D-omega-RG} 
%in a way described in ref.~\onlinecite{nphys} 
with the dimensionality $d=2$ yields the parameter values summarized in the Supplementary Table 1.

Tongue genioglossus (TG): the fitted value of the unrestricted diffusivity $D_0$ agrees well with the longitudinal eigenvalue $\lambda_1$. This already indicates that the transverse diffusion is predominantly restricted by the fiber walls, whereas the diffusion within the fibers (at short times) is approximately isotropic. 

Heart (H): since the transverse eigenvalues for the {heart} muscle exhibit the dependence \eqref{Dcum-asy} for the whole time range, for a stable fit we needed to fix one of the fit parameters; we chose to set $D_0$ for the heart to the value of the corresponding non-time-dependent eigenvalue $\lambda_1$, assuming that to correspond to the unrestricted diffusion coefficient within the fibers in analogy with the above case of the tongue muscle.

From the surface-to-volume ratio we estimate the typical ``cell diameter'' $a\simeq 2d/(S/V)$ which agrees with histological values for the actual muscle fiber diameters in both kinds of muscle ($a=20-40\,\mu$m for tongue and $a=6-12\,\mu$m for heart, see ref.~\onlinecite{kim-mrm2005} for references), yielding the heart fibers to be much narrower.
The diameter values are closer to histology than those determined in ref.~\onlinecite{kim-mrm2005} using a fit to an empirical two-compartment model.

Furthermore, our model \eqref{D-omega-RG} allows us to determine the membrane permeability values $\kappa \sim 10^{-3}\,$cm/s.
The $\kappa$ value for the tongue muscle agrees well with that expected for cell plasma membranes in eukaryotic cells  \cite{farinas}. Unfortunately, there is no ``gold standard'' noninvasive method to determine membrane permeability, which makes precise validation of the values of $\kappa$ currently unfeasible. However, qualitatively and quantitatively, these values are meaningful. 
In particular, the apparent permeability in the heart is almost as large as that of a red blood cell membrane  \cite{benga}, a few times more permeable than that of the tongue fibers. Such an elevated value is to be expected, since in the heart, there is an abundance of highly permeable blood capillaries of a similar diameter aligned with fibers, which is likely to increase the average permeability of all barriers.
We also note that, while our model  \cite{nphys} does not include extracellular space, its effect is arguably not crucial, as most of the water in tissues (80\% or more  \cite{sykova-nicholson-2008}) is contained inside cells. Based on the permeability values, on the measured time scales water molecules are able to enter and exit muscle fibers, rather than being confined within them or within the extracellular space, further justifying the use of the relatively simple random-membrane geometry of \eqref{D-omega-RG}.

%%%%%%%%%%%%%%%%%%%%%%%%%%%%%%%%%%%%%%%%%%%%%%%
\section{Origins of time-dependent diffusion in brain}

Here we demonstrate how the low-frequency dispersion \eqref{Dw} with $\alpha=1/2$ allows one to radically narrow down the scope of the plausible scenarios of the structural or functional changes in ischemic stroke.
We note that the relatively high frequencies (kHz) for dMRI in Fig.~\ref{fig:gore} appear to be low from the point of the brain microstructure at the $\mu$m scale, with the clinical dMRI accessing $D_{\infty}\equiv \D(\w)|_{\w=0}$ only.

{\bf (i)} {\it Active or passive transport?}
%The observation that the dispersion \eqref{Dw} stays qualitatively the same after injury is incompatible with active water transport as a defining mechanism for water motion in live cells. 
While active axonal transport, cytoplasmic streaming and microcirculation have been discussed as possible reasons for why diffusion might be enhanced in normal tissue relative to postmortem (see recent ref.~\onlinecite{nevo2010} for a review), it seems unlikely that the combination of these effects alone could yield the power-law dispersion \eqref{Dw} even in a normal case. Indeed, the lack of a time scale in equation~\eqref{Dw} means that these streaming processes must happen on multiple time scales,  fine-tuned in such a way as to produce the exact power law exponent $1/2$. Even if this were the case, such fine-tuning must break down after cell death with those processes switched off, causing the dispersion to change qualitatively, which contradicts Fig.~\ref{fig:gore}b. Hence we conclude that water motion is mostly determined by the ordinary diffusion hindered by the {\it passive} restrictions, and the change in their properties after injury is only quantitative, not qualitative.

{\bf (ii)} {\it Order or disorder?}
The dispersion \eqref{Dw} is {\it non-analytic} in frequency. Hence, neither bounded motion (water confined e.g. to impermeable cells of finite volume, or to effectively disconnected pockets of the extracellular space), nor any periodic structures (e.g. periodic permeable barriers, periodic beads \cite{budde-frank2010}, or any packing with a single pronounced length scale in any dimensionality  \cite{EMT}) provide the dominant cause for the observed dispersion. Indeed, all of those cases correspond to $\alpha=\infty$. The quantity $\D(\w)$ in this case is an analytic function of $\w$, i.e. it can be Taylor-expanded for small $\w$. As the velocity autocorrelator $\langle v(t)v(0)\rangle$ is real, the measured real part  \cite{EMT,sv-og} of its Fourier transform must be an even function of $\w$. This means that its Taylor expansion at small $\w$ must start with an $\w^{2}$ term, 
%$\alpha=2$ in equation \eqref{Dw}.
$\Re \D(\w) = D_\infty + \mbox{const}\cdot \w^2 + {\cal O}(\w^4)$. 
This analytic behavior was experimentally demonstrated in ref.~\onlinecite{stepisnik2007} for porous samples with impermeable walls, and is inconsistent with Fig.~\ref{fig:gore}.
Hence, the predominant restrictions to diffusion in ref.~\onlinecite{gore2003} are {\it nonconfining and disordered}.

{\bf (iii)} {\it Which disorder class?}
The passive restrictions, while not completely ordered, are still {\it correlated in space} so as to yield the $\w^{1/2}$ behavior. This observation is crucial: in the absence of any structural correlations ($p=0$), from equation \eqref{alpha=p+d} one expects the $\w^{3/2}$ dispersion in a random $d=3$ dimensional medium. Hence, either the disorder is long-range correlated (cf. Fig.~\ref{fig:randomlines}), or the effective dimensionality is less than 3.
We also note that the nonzero value of $D_{\infty}$
rules out the ``anomalous'' diffusion  \cite{Bouchaud}, e.g. in a fractal geometry --- in other words, the structural correlations are gradually forgotten and the dynamics asymptotically becomes Markoffian.
%, for which $D_{\infty}\equiv 0$. 

The power law exponent $\alpha=1/2$ can arise due to the two remaining classes of the passively restricted diffusion:

$\bullet$ $d>1$: The extended disorder with $d_{s}=1$ in $d=2$ or $d_{s}=2$ in $d=3$, Fig.~\ref{fig:randomlines}. The permeable barriers may correspond to either plasma membranes of neurons, glial cells, and of their processes (conceptually similar to the above example of muscle fiber membranes), or the membranes surrounding intracelluar organelles, such as nuclear envelope or endoplasmic reticulum.
%We will discuss below why this is unlikely.
The analysis, based on equation \eqref{D-omega-RG}, renders this possibility unlikely based both on the corresponding length scales and the permeability values. 
%shows that their permeability values are an order of magnitude higher than what is known for the most permeable eukaryotic cells, which renders this geometry unlikely.

$\bullet$ $d=1$: Any short-ranged disorder, with the effective dimensionality $d=1$, such as the $p=0$ example in Fig.~\ref{fig:1d}. Below we will argue that this is consistent with an effectively one-dimensional water motion along locally straight narrow neurites (dendrites and axons) as well as the processes of glial cells, assuming their walls to be impermeable, with some structural disorder (e.g. beads and shafts) along the way.

{\bf (iv)} {\it Extracellular water is less important.}
The much-debated contribution of the extracellular water  \cite{Benveniste} does not contribute to the observed $\alpha=1/2$ dispersion. Indeed, its effective dimensionality, assuming no exchange with cells, would be either $d=2$ (due to tight cell packing  \cite{sykova-nicholson-2008}) or $d=3$. Any short range disorder ($p=0$) in the extracellular space would then result in $\alpha=1$ or $3/2$ correspondingly, making this contribution
%$\Re \D(\w) = D_\infty + \mbox{const}\cdot |\w|$ 
%(refs.~\onlinecite{Ernst-I,Visscher}), 
less relevant for the observed dispersion: in a superposition of the $\w^{1/2}$ contribution, and of the $|\w|^{1}$ or $\w^{3/2}$ contributions, the $\w^{1/2}$ dominates as $\w\to0$. This complements the dMRI measurements of various intracellular metabolites
 \cite{wick'95,neil-mrm1996,duong-ackerman-mrm1998,dijkhuizen'99,ackerman-neil-nbm2010}, indicating that major ischemia-related changes occur already in the intracellular space.

{\bf The case of $p=0$ and $d=1$ (neurites).} Let us now assume that the neurites are impermeable and locally straight narrow one-dimensional channels with water volume fraction $\phi_{1d}$, similar to those suggested in refs.~\onlinecite{charmed,Jespersen2007}. 
In contrast to refs.~\onlinecite{charmed,Jespersen2007}, we would not assume them to be hollow cylinders; rather, we allow structural disorder along the channels, leading to the dispersive effective one-dimensional diffusivity $\D_{1d}(\w)$ identical for each channel.
Then, the measured dispersion of diffusion in a particular direction
\be \label{D1d}
\D(\w) \simeq \phi_{1d} \D_{1d}(\w)/3 + (1-\phi_{1d}) D_e \,.
\ee 
Here the factor $1/3$ assumes approximately isotropic directional distribution of neurites in gray matter, and $D_e$ now is the effective diffusivity of water outside the neurites. We can approximately set $D_e=\mbox{const}$, as the residual dispersion in the extra-neurite space should be less singular than $\w^{1/2}$ since its dimensionality $d>1$ (cf. our discussion in (iv) above). We then focus on the one-dimensional channel dispersion $\D_{1d}(\w)$ from equation \eqref{D1d}, which
%\be \label{D1d}
%\D_{1d}(\w) = 3\lb  \D(\w) - (1-\phi_{1d}) D_e\rb /\phi_{1d}
%\ee
imposes constraints on the possible parameter values. 

Our main constraint will be on the neurite volume fraction $\phi_{1d}$.
First, since  $\Re \D_{1d}(\w)>0$, 
\be \label{De<}
(1-\phi_{1d}) D_e < \D_{\rm min}\equiv \D(0) \,,
\ee
where the long-time limit
$\D_{\rm min} \approx 0.74\,\mu{\rm m}^2/$ms in normal and $\D_{\rm min} \approx 0.5\,\mu{\rm m}^2/$ms in globally ischemic brain, Fig.~\ref{fig:gore}.
Second, the $d=1$ diffusivity $\Re \D_{1d}(\w) < D_{\rm cyt}$ cannot exceed that of water in cytoplasm, $D_{\rm cyt} < 3\,\mu{\rm m}^2/$ms; from ref.~\onlinecite{Beaulieu-1994} the axoplasm diffusivity is about 80\% of that of pure water, yielding $D_{\rm cyt} \approx 2.4\,\mu{\rm m}^2/$ms as a plausible estimate.  Thus from equations~\eqref{D1d} and \eqref{De<}, we obtain 
\be \label{phi1d>}
3(\D_{\rm max}-\D_{\rm min})/\phi_{1d} < \D_{1d}(\w)|_{\w\to\infty} \equiv D_{\rm cyt} \,,
\ee
where $\D_{\rm max}$ is the maximal measured $\Re \D(\w)$, Fig.~\ref{fig:gore}. Taking into account the measurements up to 1\,kHz of the cos waveform  \cite{gore2003}, the range is at least
$\D_{\rm max}-\D_{\rm min} \approx 0.3\,\mu$m$^{2}$/ms for both normal and globally ischemic brain,
while the diffusion length at 1\,kHz still exceeds the neurite inner diameter  of a fraction of $\mu$m, 
so that the motion within the channel remains one-dimensional.
Hence, from equation \eqref{phi1d>} it follows that in order to achieve the observed dispersion with one-dimensional neurites, their volume fraction  should be sufficiently large, $\phi_{1d}\gtrsim 0.4$. This is consistent with the neurite volume fraction $\phi_{1d}\approx 0.6$ measured with electron microscopy  \cite{chklovskii-optimization2002}.

%{\bf (v)} 
%No matter which of the disorder types discussed in (iii) is realized, 
{\bf Which physical parameter changes most with ischemia?}
First we note that the  observed 50\% {increase} of the coefficient in front of $\w^{1/2}$ under ischemia, Fig.~\ref{fig:gore}, generally signifies a relative {\it increase of the disorder}, 
$\langle (\delta D)^{2}\rangle/\Dinf^{2}$,
causing the $\w^{1/2}$ dispersion [cf. \eqref{dDw}]. 
One consequence of this observation is that this apparent disorder increase is inconsistent with another debated scenario of why $\Dinf$ drops in stroke --- a suggestion that the cytoplasm itself becomes more ``viscous'', or ``dense'', causing the decrease of the free diffusion $D_{0}$ in ischemia, and with that, of $\Dinf$.
Physically, the decrease of $D_0$ would either not affect the structural disorder or cause its relative decrease, by reducing the contrast between regions with freely diffusing water and any bottlenecks or barriers.
Indeed, as a concrete example, the prefactor of $\w^{1/2}$ in the small-$\w$ expansion of equation \eqref{D-omega-RG} decreases with the decrease of $D_{0}$ for any dimensionality $d$. 
%The decrease in disorder would lead to the decrease in the dispersive $\w^{1/2}$ contribution to $\D(\w)$. 
%This intuition is confirmed by the analysis of the formula \eqref{D-omega-RG} in any $d$, and of the results of refs.~ \cite{Ernst-I,Visscher} in $d=1$.
Hence, the signature decrease in $D_\infty$ in ischemia cannot be explained by assuming that the cell cytoplasm becomes ``denser'' or ``more viscous'' after injury. This again hints at {\it the major changes being structural rather than molecular}.

A crude estimate for the change in the one-dimensional disorder after ischemia could be made using our model \cite{nphys} of randomly placed barriers, equation \eqref{D-omega-RG}, in $d=1$ dimension. For that, we assume that, on the length scales longer than a few $\mu$m, the narrow shafts between beads and spines act as effective barriers. Their effective one-dimensional ``permeability'' would be smaller for either narrower shafts, or thicker beads, and grows with the ratio between the diameters of the shafts and the beads.  

The difficulty here is that the empirical dependence in Fig.~\ref{fig:gore} allows one to obtain only two parameters, as the measurement \cite{gore2003} does not extend towards high enough $\w$ for which $\D(\w)$ saturates. Hence, as above, we need to ensure that the parameters such as $\phi_{1d}$ and $D_{e}$ stay within the above bounds.
With that in mind,  we fit the real part of  equation \eqref{D-omega-RG} with $d=1$ to $\D_{1d}(\w)$ from \eqref{D1d}.
%, with the results shown in Fig.~\ref{fig:gore-1d}b. 
Choosing $\phi_{1d}$ and $D_{e}$ within the above bounds (so that, e.g., $\Re \D_{1d}(\w)<D_{\rm cyt}$), we obtain reasonable fit results by setting $\phi_{1d}=0.7$ for both before and after global ischemia, this fraction being slightly greater than that observed in the neuropil  \cite{chklovskii-optimization2002}. This requires the diffusivity $D_e$ outside the neurites to decrease from about 1.9 before to $1.2\,\mu{\rm m}^2/$ms after ischemia. This may be explained by the effect of cell swelling making the geometry more ``tortuous''  \cite{sykova-nicholson-2008} which, as argued in (iv) above, does not affect the $\w^{1/2}$ dispersion.
We find that the effective barrier permeability $\kappa$ drops from about $0.5\,\mu$m/ms before to $0.2\,\mu$m/ms after ischemia, while the distance between ``barriers'' increases from $a=2$ to $3\,\mu$m.
The permeability reduction is consistent with the more pronounced beads (stronger contrast between bead and shaft diameters) in injured dendrites, and the increase in the distance is consistent with the disappearance of spines  \cite{zhang-murphy-2005}, such that the disorder correlation length increases.

}

%\setcounter{NAT@ctr}{40}
%\setcounter{enumi}{40}
%\begin{thebibliography}{99}
%\setcounter{enumi}{40}
%
%\bibitem{latour-jmra--}
%Latour, L. L., Mitra, P. P., Kleinberg, R. L. \& Sotak, C. H. 
%Time-dependent diffusion coefficient of fluids in porous media as a probe of surface-to-volume ratio.
%{\it J. Magn. Reson. Ser A} {\bf 101}, 342--346 (1993).
%%Latour LL, Mitra PP, Kleinberg RL, and Sotak CH (1993)
%%Time-dependent diffusion coefficient of fluids in porous media as a probe of surface-to-volume ratio.
%%{\it J Magn Reson Ser A} {101}:342-346.
%
%\end{thebibliography}

\end{document}